\documentclass[prd,aps,nofootinbib,notitlepage,showpacs,preprintnumbers]{revtex4-1}
\usepackage{graphicx,epsf,amsmath,amsfonts,amssymb,amsbsy}
\usepackage{epsfig}
\newcommand{\ds}{\displaystyle}
\newcommand{\vev}[1]{\langle#1\rangle}
\newcommand{\mat}{\left ( \begin{array}}
\newcommand{\emat}{\end{array} \right )}
\newcommand{\vect}{\left ( \begin{array}{c}}
\newcommand{\evect}{\end{array} \right )}

\begin{document}

\title{Competition and duality correspondence between inhomogeneous
fermion-antifermion and fermion-fermion condensations in the NJL$_2$ model }

\author{D. Ebert $^{1)}$, T.G. Khunjua $^{2)}$, K.G. Klimenko $^{3,4)}$, and V.Ch. Zhukovsky $^{2)}$}
\vspace{1cm}

\affiliation{$^{1)}$ Institute of Physics, Humboldt-University Berlin, 12489
Berlin, Germany} \affiliation{$^{2)}$ Faculty of Physics, Moscow State
University, 119991, Moscow, Russia} \affiliation{$^{3)}$ Institute for High
Energy Physics, 142281, Protvino, Moscow Region, Russia} \affiliation{$^{4)}$
Dubna International University (Protvino branch), 142281, Protvino, Moscow
Region, Russia}

\begin{abstract}
We investigate the possibility of spatially homogeneous and inhomogeneous
chiral fermion-antifermion condensation and superconducting fermion-fermion
pairing in the (1+1)-dimensional model by Chodos {\it et al.} [ Phys. Rev. D 61, 045011
(2000)] generalized to continuous chiral invariance. The consideration is
performed at nonzero values of temperature $T$, electric charge chemical
potential $\mu$ and chiral charge chemical potential $\mu_5$. It is shown that
at $G_1<G_2$, where $G_1$ and $G_2$ are the coupling constants in the
fermion-antifermion and fermion-fermion channels, the $(\mu,\mu_5)$-phase
structure of the model is in a one-to-one correspondence with the phase
structure at $G_1>G_2$ (called duality correspondence). Under the duality
transformation the (inhomogeneous) chiral symmetry breaking (CSB) phase is
mapped into the (inhomogeneous) superconducting (SC) phase and vice versa. If
$G_1=G_2$, then the phase structure of the model is self-dual. Nevertheless,
the degeneracy between the CSB and SC phases is possible in this case only when
there is a spatial inhomogeneity of condensates.
\end{abstract}


\maketitle

\section{Introduction}

In recent years much attention has been devoted to the investigation of dense
quark (or baryonic) matter. The interest is motivated by the possible existence
of quark matter inside compact stars or its creation in heavy ion collisions.
In many cases, as e.g. in the above-mentioned heavy ion collision experiments
the quark matter densities are not too high, so the consideration of its
properties is not possible in the framework of perturbative weak coupling QCD.
Usually, different effective theories such as the Nambu -- Jona-Lasinio (NJL)
model, $\sigma$ model etc. are more adequate in order to study the QCD and
quark matter phase diagram in this case. A variety of spatially nonuniform
(inhomogeneous) quark matter phases related to chiral symmetry breaking, color
superconductivity, and charged pion condensation phenomenon etc. (see, e.g.,
\cite{3+1,nakano,Tatsumi:2014cea,osipov,nickel,maedan,Heinz:2013eu,pisarski,miransky,zfk,incera,Anglani:2013gfu}
and references therein) was predicted in the framework of NJL-like models at
rather low values of temperature and baryon density. 
(A recent interesting review on current model results for inhomogeneous
phases in (3+1)-dimensional systems is presented in
\cite{buballa}.)

Moreover, the phenomenon of spatially nonuniform quark pairing was also
intensively investigated within different (1+1)-dimensional models which can
mimic qualitatively the QCD phase diagram. In this connection, it is necessary
to mention  Gross-Neveu (GN) type models with four-fermion interactions,
symmetrical with respect to the discrete or continuous chiral transformations
(in the last case we shall use for such models the notation NJL$_2$) and
extended by baryon and isospin chemical potentials. In the framework of these
models both the inhomogeneous chiral \cite{thies,thies2,basar} and charged pion
condensation phenomena were considered \cite{gubina,gubina2}. (In order to
overcome the prohibition on the spontaneous breaking of a continuous symmetry
in (1+1) dimensions, the consideration is usually performed in the limit of
large $N$, where $N$ is the number of quark multiplets.) 
Inhomogeneous phases in some one-dimensional organic materials and nonrelativistic Fermi gases were recently studied, correspondingly, in \cite{caldas} and \cite{Roscher} in terms of (1+1)-dimensional theories with four-fermion interaction.

Among a variety of GN-type models, there is one which describes competition
between quark-antiquark (or chiral) and quark-quark (or superconducting)
pairing at nonzero temperature $T$ and quark number chemical potential $\mu$
\cite{chodos}. Originally, the model was called for to shed new light on the
color superconductivity phenomenon in real dense quark matter. Moreover, in
\cite{chodos} the consideration is performed in the supposition that chiral and
superconducting condensates are spatially homogeneous. In this case it was
shown there that if $G_1>G_2$, where $G_1$ and $G_2$ are the coupling constants
in the chiral and superconducting channels, correspondingly, then at rather
high values of quark number chemical potential $\mu$ the superconducting phase
is realized in the system.

Since in the true ground state of any system with nonzero density the
condensates could be inhomogeneous, the aim of the present paper is to
investigate such a possibility. Namely, we shall study the phase structure of
the extended model \cite{chodos} (which is symmetric with respect to continuous
$U_A(1)$ chiral group), assuming that both quark-antiquark and quark-quark
condensates might have a spatial inhomogeneity in the form of the Fulde-Ferrel
single plane wave ansatz \cite{ff}, for simplicity. Moreover, in addition to
the particle (or quark) number chemical potential $\mu$, we also introduce into
consideration the chiral charge chemical potential $\mu_5$, which is
responsible for a nonzero chiral charge density $n_5$, i.e. to a nonzero
imbalance between densities of left- and right-handed quarks (fermions). In
literature, there are some investigations of QCD-like effective theories with
$\mu$ and $\mu_5$ chemical potentials, related to a possible parity breaking
phenomena of dense quark gluon plasma (see, e.g., \cite{andrianov,andrianov2}). Moreover, it was recently established that in heavy ion collision experiments a nonzero chiral charge density $n_5$ can be induced, leading to the so-called chiral magnetic effect \cite{ruggieri,huang}. So, we hope that studying the above- mentioned (1+1)-dimensional NJL model with two chemical potentials, $\mu$ and $\mu_5$,  one can shed new light on the new phenomena of the dense baryonic matter.

The paper is organized as follows. In Sec. II the duality property of the
model is established. It means that there is a correspondence between
properties (phase structure) of the model at $G_1<G_2$ and $G_1>G_2$. After
obtaining the thermodynamic potential (TDP), we will first investigate it in
the next Sec. III under the supposition that both superconducting and chiral
condensates are spatially homogeneous. In this section a rather rich
$(\mu,\mu_5)$-phase structure of the model is established at $G_1>G_2$. In
addition, we will show here that there is an invariance of the TDP with respect
to a duality transformation (when $G_1\leftrightarrow G_2$,
$\mu\leftrightarrow\mu_5$ and superconductivity $\leftrightarrow$ chiral
symmetry breaking). As a result, the $(\mu,\mu_5)$-phase structure of the model
at $G_1<G_2$ is a dual mapping of the phase portrait at $G_1>G_2$. In Sec. IV the phase structure of the model is investigated in the assumption that both
condensates might be spatially inhomogeneous. Then at $G_1>G_2$ the chiral
density wave phase is realized for arbitrary values of $\mu\ne 0$ and $\mu_5\ne
0$. On the other hand, at $G_1<G_2$ there is an inhomogeneous superconducting
phase in the whole $(\mu,\mu_5)$ plane. Note, that there is a dual
correspondence between these phases. Finally, Sec. V presents a summary and
some concluding remarks. The discussion of some technical problems are
relegated to four Appendixes.

\section{ The model and its thermodynamic potential}
\label{effaction}

\subsection{The duality property of the model }

Our investigation is based on a (1+1)-dimensional NJL-type model with massless
fermions belonging to a fundamental multiplet of the $O(N)$ flavor group. Its
Lagrangian describes the interaction in the fermion--antifermion and
scalar fermion-fermion channels,
\begin{eqnarray}
 L=\sum_{k=1}^{N}\bar \psi_k\Big [\gamma^\nu i\partial_\nu
+\mu\gamma^0+\mu_5\gamma^0\gamma^5\Big ]\psi_k&+& \frac {G_1}N\left[\left
(\sum_{k=1}^{N}\bar \psi_k\psi_k\right )^2+\left (\sum_{k=1}^{N}\bar \psi_k
i\gamma^5\psi_k\right )^2\right ]\nonumber\\&&~~~~~~~~~~~~~~~+\frac {G_2}N\left
(\sum_{k=1}^{N} \psi_k^T\epsilon\psi_k\right )\left (\sum_{j=1}^{N}\bar
\psi_j\epsilon\bar\psi_j^T\right ), \label{1}
\end{eqnarray}
where $\mu$ is a fermion number chemical potential (conjugated to a fermion, or
electric charge, number density) and $\mu_5$ is an axial chemical potential
conjugated to a nonzero density of chiral charge $n_5=n_R-n_L$, which
represents an imbalance in densities of the right- and left-handed fermions
\cite{ruggieri}. As it is noted above, all fermion fields $\psi_k$
($k=1,...,N$) form a fundamental multiplet of the $O(N)$ group. Moreover, each
field $\psi_k$ is a two-component Dirac spinor (the symbol $T$ denotes the
transposition operation). The quantities $\gamma^\nu$ ($\nu =0,1$), $\gamma^5$,
and $\epsilon$ in (1) are matrices in the two-dimensional spinor space,
\begin{equation}
\begin{split}
\gamma^0=\begin{pmatrix}
0&1\\
1&0\\
\end{pmatrix};\qquad
\gamma^1=\begin{pmatrix}
0&-1\\
1&0\\
\end{pmatrix}\equiv -\epsilon;\qquad
\gamma^5=\gamma^0\gamma^1=
\begin{pmatrix}
1&0\\
0&{-1}\\
\end{pmatrix}.
\end{split}\label{2}
\end{equation}
It follows from (\ref{2}) that $\mu_5\gamma^0\gamma^5=\mu_5\gamma^1$. Clearly,
the Lagrangian $L$ is invariant under transformations from the internal $O(N)$
group, which is introduced here in order to make it possible to perform all the
calculations in the framework of the nonperturbative large-$N$ expansion
method. Physically more interesting is that the model (1) is invariant under
transformations from the $U_V(1)\times U_A(1)$ group, where $U_V(1)$ is the
fermion number conservation group, $\psi_k\to\exp (i\alpha)\psi_k$
($k=1,...,N$), and $U_A(1)$ is the group of continuous chiral transformations,
$\psi_k\to\exp (i\alpha'\gamma^5)\psi_k$ ($k=1,...,N$). \footnote{Earlier in
\cite{chodos} a similar model symmetric under discrete $\gamma^5$ chiral
transformation was investigated. However,  only the possibility for the
spatially homogeneous chiral and difermion condensates was considered there. In
our paper, the invariance  of the model considered by Chodos {\it et al.} \cite{chodos}
is generalized to the case of continuous chiral symmetry in order to study the
inhomogeneous chiral condensates in the form of chiral spirals (or chiral
density waves).} The linearized version of Lagrangian (\ref{1}) that contains
auxiliary scalar bosonic fields $\sigma (x)$, $\pi (x)$, $\Delta(x)$,
$\Delta^{*}(x)$ has the following form:
\begin{eqnarray}
{\cal L}\ds\equiv {\cal L}(G_1,G_2;\mu,\mu_5)&=&\bar\psi_k\Big [\gamma^\nu
i\partial_\nu
+\mu\gamma^0+\mu_5\gamma^1 -\sigma -i\gamma^5\pi\Big ]\psi_k\nonumber\\
&-&\frac{N}{4G_1}(\sigma^2+\pi^2) -\frac N{4G_2}\Delta^{*}\Delta-
 \frac{\Delta^{*}}{2}[\psi_k^T\epsilon\psi_k]
-\frac{\Delta}{2}[\bar\psi_k \epsilon\bar\psi_k^T]. \label{3}
\end{eqnarray}
(Here and in what follows, summation over repeated indices $k=1,...,N$ is
implied.) Clearly, the Lagrangians (\ref{1}) and (\ref{3}) are equivalent, as
can be seen by using the Euler-Lagrange equations of motion for scalar bosonic
fields which take the form
\begin{eqnarray}
\sigma (x)=-2\frac {G_1}N(\bar\psi_k\psi_k),~~\pi (x)=-2\frac {G_1}N(\bar\psi_k
i\gamma^5\psi_k),~~ \Delta(x)=-2\frac {G_2}N(\psi_k^T\epsilon\psi_k),~~
\Delta^{*}(x)=-2\frac {G_2}N(\bar\psi_k \epsilon\bar\psi_k^T). \label{4}
\end{eqnarray}
One can easily see from (\ref{4}) that the (neutral) fields $\sigma(x)$ and
$\pi(x)$ are real quantities, i.e. $(\sigma(x))^\dagger=\sigma(x)$,
$(\pi(x))^\dagger=\pi(x)$ (the superscript symbol $\dagger$ denotes the
Hermitian conjugation), but the (charged) difermion scalar fields $\Delta(x)$
and $\Delta^*(x)$ are Hermitian conjugated complex quantities, so
$(\Delta(x))^\dagger= \Delta^{*}(x)$ and vice versa. Clearly, all the fields
(\ref{4}) are singlets with respect to the $O(N)$ group. \footnote{Note that
the $\Delta (x)$ field is a flavor $O(N)$ singlet, since the representations of
this group are real.} If the scalar difermion field $\Delta(x)$ has a nonzero
ground state expectation value, i.e.\  $\vev{\Delta(x)}\ne 0$, the Abelian
fermion number $U_V(1)$ symmetry of the model is spontaneously broken down.
However, if $\vev{\sigma (x)}\ne 0$ then the continuous chiral symmetry of the
model is spontaneously broken.

Before studying the thermodynamics of the model, we want first of all to
consider its duality property. To this end, it is very useful to form an
infinite set $\boldsymbol {\cal F}$ composed of all Lagrangians ${\cal
L}(G_1,G_2;\mu,\mu_5)$ (\ref{3}) when the free model parameters $G_1,G_2,\mu$
and $\mu_5$  take arbitrary admissible values, i.e. ${\cal
L}(G_1,G_2;\mu,\mu_5)\in \boldsymbol {\cal F}$ at arbitrary fixed values of
coupling constants $G_1>0, G_2>0$ and chemical potentials $\mu, \mu_5$. Then,
let us perform in (\ref{3}) the so-called Pauli-Gursey transformation of
spinor fields \cite{pauli}, accompanied with corresponding simultaneous
transformations of auxiliary scalar fields (\ref{4}),
 \begin{eqnarray}
\psi_k (x)\longrightarrow \frac 12 (1-\gamma^5)\psi_k (x)+\frac 12
(1+\gamma^5)\epsilon\bar\psi^T_k(x);~~ \sigma (x)\rightleftarrows
\frac{\Delta(x)+\Delta^{*}(x)}{2};~~\pi (x)\rightleftarrows
\frac{\Delta(x)-\Delta^{*}(x)}{2i}. \label{04}
\end{eqnarray}
Taking into account that all spinor fields anticommute with each other, it is
easy to see that under the action of the transformations (\ref{04}) each
element (auxiliary Lagrangian) ${\cal L}(G_1,G_2;\mu,\mu_5)$ of the set
$\boldsymbol {\cal F}$ is transformed into another element of the set
$\boldsymbol {\cal F}$ according to the following rule
 \begin{eqnarray}
{\cal L}(G_1,G_2;\mu,\mu_5) \longrightarrow {\cal L}(G_2,G_1;-\mu_5,-\mu)\in
\boldsymbol {\cal F}, \label{004}
\end{eqnarray}
i.e. the set $\boldsymbol {\cal F}$ is invariant under the field
transformations (\ref{04}). Owing to the relation (\ref{004}) there is a
connection between properties of the model when free model parameters
$G_1,G_2,\mu$ and $\mu_5$ vary in different regions. Due to this reason, we
will call the relation (\ref{004}) the duality property of the model.

\subsection{The thermodynamic potential at $T=0$}

We begin an investigation of a phase structure of the four-fermion model (1)
using the equivalent semibosonized Lagrangian (\ref{3}). In the leading order
of the large-$N$ approximation, the effective action ${\cal S}_{\rm
{eff}}(\sigma,\pi,\Delta,\Delta^{*})$ of the considered model is expressed by
means of the path integral over fermion fields:
$$
\exp(i {\cal S}_{\rm {eff}}(\sigma,\pi,\Delta,\Delta^{*}))=
\int\prod_{l=1}^{N}[d\bar\psi_l][d\psi_l]\exp\Bigl(i\int {\cal
 L}\,d^2 x\Bigr),
$$
where
\begin{eqnarray}
&&{\cal S}_{\rm {eff}} (\sigma,\pi,\Delta,\Delta^{*}) =-\int d^2x\left
[\frac{N}{4G_1}(\sigma^2(x)+\pi^2(x))+ \frac{N}{4G_2}\Delta
(x)\Delta^{*}(x)\right ]+ \widetilde {\cal S}_{\rm {eff}}. \label{5}
\end{eqnarray}
The fermion contribution to the effective action, i.e.\  the term $\widetilde
{\cal S}_{\rm {eff}}$ in (\ref{5}), is given by
\begin{equation}
\exp(i\widetilde {\cal S}_{\rm
{eff}})=\int\prod_{l=1}^{N}[d\bar\psi_l][d\psi_l]\exp\Bigl\{i\int\Big
[\bar\psi_k(\gamma^\nu i\partial_\nu
+\mu\gamma^0+\mu_5\gamma^1-\sigma-i\gamma^5\pi)\psi_k -
 \frac{\Delta^{*}}{2}(\psi_k^T\epsilon\psi_k)
-\frac{\Delta}{2}(\bar\psi_k \epsilon\bar\psi_k^T)\Big ]d^2 x\Bigr\}. \label{6}
\end{equation}
The ground state expectation values $\vev{\sigma(x)}$, $\vev{\pi(x)}$, etc. of
the composite bosonic fields are determined by the saddle point equations,
\begin{eqnarray}
\frac{\delta {\cal S}_{\rm {eff}}}{\delta\sigma (x)}=0,~~~~~ \frac{\delta {\cal
S}_{\rm {eff}}}{\delta\pi (x)}=0,~~~~~ \frac{\delta {\cal S}_{\rm
{eff}}}{\delta\Delta (x)}=0,~~~~~ \frac{\delta {\cal S}_{\rm
{eff}}}{\delta\Delta^* (x)}=0. \label{7}
\end{eqnarray}
In vacuum, i.e. in the state corresponding to an empty space with zero particle
density and zero values of the chemical potentials $\mu$ and $\mu_5$, the above-mentioned quantities $\vev{\sigma(x)}$, etc. do not depend on space
coordinates. However, in a dense medium, when $\mu\ne 0$ and/or $\mu_5\ne 0$,
the ground state expectation values of bosonic fields (\ref{4}) might have a
nontrivial dependence on the spatial coordinate $x$. In particular, in this
paper we will use the following ansatz:
\begin{eqnarray}
\vev{\sigma(x)}=M\cos (2bx),~~~\vev{\pi(x)}=M\sin
(2bx),~~~\vev{\Delta(x)}=\Delta\exp(2ib'x),~~~
\vev{\Delta^*(x)}=\Delta\exp(-2ib'x), \label{8}
\end{eqnarray}
where $M,b,b'$ and $\Delta$ are real constant quantities. (It means that we
suppose for $\vev{\sigma(x)}$ and $\vev{\pi(x)}$ the chiral spiral (or chiral
density wave) ansatz, and the Fulde-Ferrel \cite{ff} single plane wave ansatz
for difermion condensates.) In fact, they are coordinates of the global minimum
point of the thermodynamic potential $\Omega (M,b,b',\Delta)$.
\footnote{Here and in what follows we will use a conventional notation "global"
minimum in the sense that among all our numerically found local minima the
thermodynamical potential takes in their case the lowest value. This does not
exclude the possibility that there exist other inhomogeneous condensates,
different from (\ref{8}), which lead to ground states with even lower values of
the TDP.} In the leading order of the large $N$-expansion it is defined by the
following expression:
\begin{equation*}
\int d^2x \Omega (M,b,b',\Delta)=-\frac{1}{N}{\cal S}_{\rm
{eff}}\{\sigma(x),\pi(x),\Delta (x),\Delta^*(x)\}\Big|_{\sigma
(x)=\vev{\sigma(x)},\pi(x)=\vev{\pi (x)},...} ,
\end{equation*}
which gives
\begin{eqnarray}
\int d^2x\Omega (M,b,b',\Delta)\,\,&=&\,\,\int d^2x\left
(\frac{M^2}{4G_1}+\frac{\Delta^2}{4G_2}\right )+\frac{i}{N}\ln\left (
\int\prod_{l=1}^{N}[d\bar\psi_l][d\psi_l]\exp\Big (i\int d^2 x\Big [\bar\psi_k
{\cal D}\psi_k\right.\nonumber\\&& \left.-
\frac{\Delta\exp(-2ib'x)}{2}(\psi_k^T\epsilon\psi_k)
-\frac{\Delta\exp(2ib'x)}{2}(\bar\psi_k \epsilon\bar\psi_k^T)\Big ] \Big
)\right ), \label{9}
\end{eqnarray}
where ${\cal D}=\gamma^\rho i\partial_\rho +\mu\gamma^0+\mu_5\gamma^1
-M\exp(2i\gamma^5bx)$. 
In principle, one way to evaluate the path integral in (\ref{9}) is to
extend the technique of the paper \cite{basar}, where a more simple
model with single quark-antiquark   
channel of interaction was investigated, to the case under
consideration, i.e. to the GN model (1) with additional
superconducting interaction of quarks. The  rigorous method of
\cite{basar} is based on finding the resolvent function corresponding
to the Hamiltonian of the system. However, technically it is very
difficult to use this approach in the framework of the model (1). So,
in order to simplify the problem we first perform in (\ref{9})
Weinberg (or chiral) transformation of spinor fields \cite{weinberg},
$q_k=\exp[i(\gamma^5b-b')x]\psi_k$ and $\bar q_k=
\bar\psi_k\exp[i(\gamma^5b+b')x]$. Since Weinberg transformation of
fermion fields does not change the path integral measure in (\ref{9}),
\footnote{Strictly speaking, performing Weinberg transformation of
  fermion fields in (\ref{9}), one can obtain in the path integral
  measure a factor, which however does not depend on the dynamical
  variables $M$, $\Delta$, $b$, and $b'$. Hence, we ignore this
  unessential factor in the following calculations. Note that only in the
  case when there is an interaction between spinor and gauge fields
  there might appear a nontrivial, i.e. dependent on dynamical
  variables, path integral measure, generated by Weinberg
  transformation of spinors. This unobvious fact follows from the
  investigations by Fujikawa \cite{fujikawa}.\label{fujik}} we see
that the system is reduced by the Weinberg transformation from a
spatially modulated to a uniform one; i.e. we obtain the following
expression  for the thermodynamic potential: 
\begin{eqnarray}
\int d^2x\Omega (M,b,b',\Delta)&=&\int d^2x\left
(\frac{M^2}{4G_1}+\frac{\Delta^2}{4G_2}\right ) \nonumber\\
&+&\frac{i}{N}\ln\left ( \int\prod_{l=1}^{N}[d\bar q_l][d q_l]\exp\Big (i\int
d^2 x\Big
[\bar q_k  D q_k
-\frac{\Delta}{2}(q_k^T\epsilon q_k) -\frac{\Delta}{2}(\bar q_k \epsilon\bar
q_k^T)\Big ] \Big )\right ) \label{10},
\end{eqnarray}
where
\begin{equation}
D=\gamma^\nu i\partial_\nu +(\mu-b)\gamma^0-M+\gamma^1(\mu_5-b'). \label{110}
\end{equation}
The path integration in the expression (\ref{10}) is evaluated in Appendix
\ref{ApA} \footnote{In Appendix \ref{ApA} we consider for simplicity the case
$N=1$; however the procedure is easily generalized to the case with $N>1$.}
(see also \cite{kzz} for similar integrals), so we have for the TDP
\begin{eqnarray}
\Omega (M,b,b',\Delta)\equiv\Omega^{un} (M,b,b',\Delta)&=&
\frac{M^2}{4G_1}+\frac{\Delta^2}{4G_2}
+\frac{i}{2}\int\frac{d^2p}{(2\pi)^2}\ln\Big [\lambda_1(p)\lambda_2(p)\Big ],
\label{11}
\end{eqnarray}
where $\lambda_{1,2}(p)$ are presented in (\ref{A8}) and superscript ``un``
denotes the unrenormalized quantity. Note, the TDP (\ref{11}) describes
thermodynamics of the model at zero temperature $T$. In the following we will
study the behavior of the global minimum point of this TDP as a function of
dynamical variables $M,b,b',\Delta$ vs the external parameters $\mu$ and
$\mu_5$ in two qualitatively different cases: (i) the case of homogeneous
condensates, i.e. when in (\ref{8}) and (\ref{11}) both $b$ and $b'$ are
supposed from the very beginning, without any proof, to be zero, and (ii) the
case of spatially inhomogeneous condensates, i.e. when the quantities $b$ and
$b'$ are defined dynamically by the gap equations of the TDP (\ref{11}).
Moreover, the influence of temperature $T$ on the phase structure is also taken
into account.

\section{The homogeneous case of the ansatz (\ref{8}) for condensates: $b=0$ and $b'=0$}

\subsection{Dual invariance of the TDP}

In the present section we suppose that all the condensates are spatially
homogeneous, i.e. we put in the ansatz (\ref{8}) and in the TDP (\ref{11})
$b\equiv 0$ and $b'\equiv 0$. So, the TDP is considered {\it a priori} as a function
of only two variables, $M$ and $\Delta$ ($\mu$ and $\mu_5$ are treated as
external parameters). Note that the subject and results of the section are largely
preparatory for considering the main purpose of the paper, i.e. to clarify (see
the next section) a genuine ground state structure of the model in the
framework of the inhomogeneous ansatz (\ref{8}) for condensates.

Taking into account the expressions (\ref{A8}) for $\lambda_{1,2}(p)$, we
obtain the unrenormalized TDP in this case:
\begin{eqnarray}
\Omega^{un} (M,\Delta)&=& \frac{M^2}{4G_1}+\frac{\Delta^2}{4G_2}
+\frac{i}{2}\int\frac{d^2p}{(2\pi)^2}\ln\Big [\det\bar B(p)\Big ], \label{25}
\end{eqnarray}
where
\begin{eqnarray}
\det\bar B(p)=\lambda_1(p)\lambda_2(p)\Big|_{b=0,b'=0}&=&\Delta^4-2\Delta^2(p_0^2-p_1^2+M^2+\mu_5^2-\mu^2)\nonumber\\
&+&\big (M^2+(p_1-\mu_5)^2-(p_0-\mu)^2\big )\big (M^2+(p_1+\mu_5)^2-
(p_0+\mu)^2\big ). \label{26}
\end{eqnarray}
Expanding the right-hand side of (\ref{26}) in powers of $M$, one can obtain an
equivalent expression for $\det\bar B(p)$. Namely,
\begin{eqnarray}
\det\bar B(p)=M^4-2M^2(p_0^2-p_1^2&+&\Delta^2+\mu^2-\mu_5^2)\nonumber\\
&+&\big (\Delta^2+(p_1-\mu)^2-(p_0-\mu_5)^2\big )\big
(\Delta^2+(p_1+\mu)^2-(p_0+\mu_5)^2\big ).\label{27}
\end{eqnarray}
We would like to stress once more that there is an identical equality between
the expressions (\ref{26}) and (\ref{27}).

 Obviously, the function $\Omega^{un} (M,\Delta)$
(\ref{25}) is symmetric with respect to the  transformations $M\to-M$ and/or
$\Delta\to -\Delta$. Moreover, it is invariant under the transformations
$\mu_5\to-\mu_5$ and/or $\mu\to-\mu$. \footnote{Indeed, if simultaneously with
$\mu_5\to-\mu_5$ and/or $\mu\to-\mu$ transformations we perform in the integral
(\ref{25}) the following change of variables, $p_1\to -p_1$ and/or $p_0\to
-p_0$, then one can easily see that the expression (\ref{26}) remains intact.}
Hence, without loss of generality, we restrict ourselves by the constraints:
$M\ge 0$, $\Delta\ge 0$, $\mu\ge 0$, and $\mu_5\ge 0$. However, there is one
more discrete transformation of the TDP (\ref{25}), which leaves it invariant.
It follows from a comparison between (\ref{26}) and (\ref{27}). Indeed, if in
(\ref{26}) for $\det\bar B(p)$ the transformations $\mu\leftrightarrow\mu_5$
and $M\leftrightarrow\Delta$ are performed simultaneously, then the expression
(\ref{27}) will be obtained, which is equal to the original expression
(\ref{26}) for $\det\bar B(p)$. So the TDP (\ref{25}) is invariant with respect
to the following duality transformation $D$:
\begin{eqnarray}
D:~~G_1\longleftrightarrow G_2,~~\mu\longleftrightarrow\mu_5,~~
M\longleftrightarrow\Delta. \label{28}
\end{eqnarray}
Taking into account that the TDP (\ref{25}) is symmetric with respect to
$\mu_5\to-\mu_5$ and/or $\mu\to-\mu$, it is possible to conclude that the dual
invariance $D$ of the TDP (\ref{25}) is a particular realization of the dual
property (\ref{004}) of the initial model.  Suppose now that at some fixed
particular values of the model parameters, i.e. at $G_1=A$, $G_2=B$ and
$\mu=\alpha$, $\mu_5=\beta$, the global minimum point of the TDP lies at the point
$(M=M_0,\Delta=\Delta_0)$. Then it follows from the dual invariance $D$
(\ref{28}) of the TDP that the permutation of the coupling constant and
chemical potential values (i.e. at $G_1=B$, $G_2=A$ and $\mu=\beta$,  $\mu_5=\alpha$)
moves the global minimum point of the TDP to the point
$(M=\Delta_0,\Delta=M_0)$. In particular, if in the original model with
$G_1=A$, $G_2=0$ and $\mu=\alpha$, $\mu_5=0$ the global minimum point of the TDP lies
at the point $(M=M_0,\Delta=0)$ (as a result, in this case the continuous
chiral symmetry $U_A(1)$ is spontaneously broken down), then in the model with
$G_1=0$, $G_2=A$ and $\mu=0$, $\mu_5=\alpha$ the global minimum point of the TDP lies
at the point $(M=0,\Delta=M_0)$ and the symmetry $U_V(1)$ is spontaneously
broken. The duality correspondence between these two particular cases of the
original model (1) was discussed in \cite{thies1}. (Even earlier, a special
case with $\mu=\mu_5=0$ of the duality between chiral symmetry breaking and
superconductivity phenomena was considered in the framework of the simplest
two-dimensional Gross-Neveu model \cite{Ojima:1977cg,Vasiliev:1995qp}.) Hence,
a knowledge of a phase structure of the model (1) at $G_1<G_2$ is sufficient to
construct, by applying the duality transformation $D$ (\ref{28}), the phase
structure at $G_1>G_2$; i.e. in the model under consideration there is a
duality correspondence between chiral symmetry breaking and superconducting
phases.

 To investigate the TDP (\ref{25}) it is necessary to renormalize it.

\subsection{The vacuum case: $\mu=0$, $\mu_5=0$}

First of all we will consider the renormalization procedure and the phase
structure of the model in the vacuum case, i.e. when $\mu=0,\mu_5=0$. Putting
$\mu=0$ and $\mu_5=0$ in (\ref{25}), we have in this case the following
expression for the unrenormalized effective potential $V_0^{un}(M,\Delta)$ (in
vacuum TDP is usually called an effective potential):
\begin{eqnarray}
V_0^{un} (M,\Delta)&=& \frac{M^2}{4G_1}+\frac{\Delta^2}{4G_2}
+\frac{i}{2}\int\frac{d^2p}{(2\pi)^2}\ln\Big [\left (p_0^2-p_1^2-
(\Delta-M)^2\right )\left (p_0^2-p_1^2-(\Delta+M )^2\right)\Big ], \label{29}
\end{eqnarray}
Integrating in (\ref{29}) over $p_0$ (see Appendix B in \cite{gubina2} for
similar integrals) and cutting the integration $p_1$ region, $|p_1|<\Lambda$,
one obtains the regularized effective potential $V_0^{reg}(M,\Delta)$:
\begin{eqnarray}
V_0^{reg} (M,\Delta)&=& \frac{M^2}{4G_1}+\frac{\Delta^2}{4G_2}-
\int_{0}^\Lambda\frac{dp_1}{2\pi}\Big
(\sqrt{p_1^2+(M+\Delta)^2}+\sqrt{p_1^2+(M-\Delta)^2}\Big ). \label{30}
\end{eqnarray}
Since this expression diverges at $\Lambda\to\infty$, it is necessary to
renormalize it, assuming that $G_1\equiv G_1(\Lambda)$ and $G_2\equiv
G_2(\Lambda)$ have appropriate $\Lambda$ dependencies. It is easy to
establish that if
 \begin{eqnarray}
\frac{1}{4G_1}\equiv\frac{1}{4G_1(\Lambda)}=\frac{1}{2\pi}\ln\frac{2\Lambda}{M_1},~~~\frac{1}{4G_2}\equiv\frac{1}{4G_2(\Lambda)}=\frac{1}{2\pi}\ln\frac{2\Lambda}{M_2},
\label{31}
\end{eqnarray}
where $M_1$ and $M_2$ are some finite and cutoff independent parameters with
dimensionality of mass, then integrating in (\ref{30}) over $p_1$ and ignoring
there an unessential term -$\Lambda^2/2\pi$ one can obtain in the  limit
$\Lambda\to\infty$ a finite and renormalization invariant expression for the
effective potential,
\begin{eqnarray}
4\pi V_0(M,\Delta)&=&
M^2\ln\frac{|M^2-\Delta^2|}{M^2_1}+\Delta^2\ln\frac{|M^2-\Delta^2|}{M^2_2}+2M\Delta\ln\left
|\frac{M+\Delta}{M-\Delta}\right |-\Delta^2-M^2.
 \label{22}
\end{eqnarray}
Now two remarks are in order. First, since  $M_1$ and $M_2$ can be considered
as  free model parameters, it is clear that the renormalization procedure of
the NJL$_2$ model (1) is accompanied by the dimensional transmutation
phenomenon. Indeed, there are two dimensionless bare coupling constants
$G_{1,2}$ in the initial unrenormalized expression (\ref{29})  for
$V_0^{un}(M,\Delta)$, whereas after renormalization the effective potential
(\ref{22}) is characterized by two dimensional, $M_1$ and $M_2$, free model
parameters. Moreover, $M_1$ and $M_2$ are renormalization invariant quantities,
i.e. they do not depend on the normalization points. (The physical sense of
$M_1$ and $M_2$ will be discussed below.) Second, the transposition
$G_1\leftrightarrow G_2$ of the bare coupling constants before renormalization
is equivalent, as it is clear from (\ref{31}), to the transposition
$M_1\leftrightarrow M_2$ after  renormalization procedure. Hence, the vacuum
effective potential $V_0(M,\Delta)$ (\ref{22}) of the model is invariant with
respect to the duality transformation (\ref{28}) which now, i.e. in vacuum,
looks like $M_1\leftrightarrow M_2$, $M\leftrightarrow\Delta$.

Note also that the effective potential $V_0(M,\Delta)$ written in the form
(\ref{22}) has a singularity at $M=\Delta$, which is really fictitious. Indeed,
the expression (\ref{22}) may be presented in an equivalent form that is more convenient
for both numerical and analytical investigations:
\begin{eqnarray}
4\pi V_0(M,\Delta)&=&\delta\Delta^2-\Delta^2-M^2+(M-\Delta)^2\ln\left
|\frac{M-\Delta}{M_1}\right |+(M+\Delta)^2\ln\left (\frac{M+\Delta}{M_1}\right
),
 \label{220}
\end{eqnarray}
where
 \begin{eqnarray}
\frac{\delta}{4\pi}\equiv\frac{1}{4G_2}-\frac{1}{4G_1}=\frac{1}{2\pi}\ln\frac{M_1}{M_2}.
\label{20}
\end{eqnarray}
The expression (\ref{220}) is now a smooth function at $M=\Delta$. As it is
clear from (\ref{220}), instead of two massive $M_1$ and $M_2$ parameters the
renormalized model can be characterized by one massive and one dimensionless
parameter $M_1$ and $\delta$, respectively. (In this case only the partial
dimensional transmutation phenomenon takes place.) Just this set of parameters,
i.e. $M_1$ and $\delta$, was used in early investigations of the initial model
(1) at $\mu_5=0$ \cite{chodos}. In spite of the fact that the dual invariance
$D$ (\ref{28}) of the effective potential in the form (\ref{220}), i.e. its
symmetry with respect to simultaneous transformations $M_1\leftrightarrow M_2$
and $M\leftrightarrow\Delta$, is not so evident as in the form (\ref{22}), in
the following we will treat the model properties  in terms of the parameters
$M_1$ and $\delta$ as well.

So, if $\delta>0$, i.e., as is easily seen from (\ref{20}) and (\ref{31}), at
$G_1>G_2$ or $M_1>M_2$, the global minimum of the effective potential
(\ref{220}) lies at the point $(M=M_1,\Delta=0)$. This means that if
interaction in the fermion-antifermion channel is greater than that in the
difermion one, then the chiral symmetry of the model is spontaneously broken
down and fermions acquire dynamically a nonzero Dirac mass, which is equal just
to the free model parameter $M_1$. Further, in order to establish the phase
structure of the model (or, equivalently, to find the global minimum point of
the function $V_0(M,\Delta)$) at $\delta<0$, i.e. at $G_1<G_2$,  we do not need
a straightforward analytical (or numerical) study of the function (\ref{220})
on the extremum. In this case it is enough to take into account the dual
invariance (\ref{28}) of the TDP (\ref{25}) (at $\mu=\mu_5=0$ it is reduced to
a symmetry of the effective potential $V_0(M,\Delta)$ with respect to
simultaneous permutations $M_1\leftrightarrow M_2$, $M\leftrightarrow\Delta$)
and conclude (see also the discussion just after (\ref{28})) that at $\delta<0$
the effective potential (\ref{220}) has a global minimum at the point
$(M=0,\Delta=\Delta_0)$, where $\Delta_0=M_2=M_1\exp(-\delta/2)$. Since in this
case only the difermion condensate, which is equal to $M_2$, is nonzero, the
fermion number $U(1)$ symmetry is spontaneously broken and the superconducting
phase is realized in the model. Hence, the parameter $M_2$ is a Majorana mass
of fermions, which appears dynamically in superconducting phase of the model.

\subsection{The case $\mu >0$, $\mu_5 >0$ and $T=0$}
\label{mu}

Taking into account the expression (\ref{B9}) (see Appendix B), in this case
the unrenormalized TDP (\ref{25}) can be presented in the following form
\begin{eqnarray}
\Omega^{un}
(M,\Delta)&=&\frac{M^2}{4G_1}+\frac{\Delta^2}{4G_2}-\int^\infty_{0}\frac{dp_1}{4\pi}\Big\{|p_{01}|+|p_{02}|+|\bar
p_{01}|+|\bar p_{02}|\Big\}, \label{32}
\end{eqnarray}
where quasiparticle and quasiantiparticle energies $p_{01}$, $p_{02}$ and
$\bar p_{01}$, $\bar p_{02}$, respectively, are presented in (\ref{B41}). It is
shown in Appendix B (see the text below  formula (\ref{B071})) how one can find
the asymptotic expansion of the integrand in (\ref{32}) at $|p_1|\to\infty$. As
a consequence of this prescription we have obtained the asymptotic expansions
(\ref{B26}) and, as a result, the following $|p_1|\to\infty$ expansion:
 \begin{eqnarray}
|p_{01}|+|p_{02}|+|\bar p_{01}|+|\bar
p_{02}|=4|p_{1}|+\frac{2(M^2+\Delta^2)}{|p_{1}|}+{\cal O}\big (1/|p_{1}|^2\big
). \label{33}
\end{eqnarray}
It means that the integral in (\ref{32}) is an ultraviolet (UV) divergent, so
we need to  renormalize the TDP $\Omega^{un} (M,\Delta)$. Using the momentum
cutoff regularization scheme, we obtain
\begin{eqnarray}
\Omega^{reg}
(M,\Delta)=\frac{M^2}{4G_1}+\frac{\Delta^2}{4G_2}&-&\int^\Lambda_{0}\frac{dp_1}{4\pi}\Big\{|p_{01}|+|p_{02}|+|\bar
p_{01}|+|\bar p_{02}|\Big\} \label{230}\\
=V_0^{reg} (M,\Delta) &-&\int_{0}^\Lambda\frac{dp_1}{4\pi}\Big
(|p_{01}|+|p_{02}|+|\bar
p_{01}|+|\bar p_{02}|\nonumber\\
&-&2\sqrt{p_1^2+(M+\Delta)^2}-2\sqrt{p_1^2+(M-\Delta)^2}\Big ), \label{23}
\end{eqnarray}
where $V_0^{reg} (M,\Delta)$ is given in (\ref{30}). Note that the leading terms of
the asymptotic expansion (\ref{33}) do not depend on $\mu$ and $\mu_5$. So the
quantity
 \begin{eqnarray}
\big (|p_{01}|+|p_{02}|+|\bar p_{01}|+|\bar p_{02}|\big )\big
|_{\mu=0,\mu_5=0}\equiv 2\sqrt{p_1^2+(M+\Delta)^2}+2\sqrt{p_1^2+(M-\Delta)^2}
\label{330}
\end{eqnarray}
has the same asymptotic expansion (\ref{33}) at $|p_1|\to\infty$. Hence, the
integral term in (\ref{23}) is a convergent one, and all UV divergences are
located in the first term $V_0^{reg} (M,\Delta)$. The  UV divergences are
eliminated, if the $\Lambda$ dependencies (\ref{31}) of the bare coupling
constants $G_1$ and $G_2$ are supposed. In this case we have from (\ref{23}) at
$\Lambda\to\infty$ the following expression for the renormalized TDP:
\begin{eqnarray}
\Omega^{ren}
(M,\Delta)&=&V_0(M,\Delta)-\int^\infty_{0}\frac{dp_1}{4\pi}\Big\{|p_{01}|+|p_{02}|+|\bar
p_{01}|+|\bar
p_{02}|-2\sqrt{p_1^2+(M+\Delta)^2}-2\sqrt{p_1^2+(M-\Delta)^2}\Big\}, \label{35}
\end{eqnarray}
where $V_0(M,\Delta)$ is the TDP (effective potential) (\ref{22})-(\ref{220})
of the model at $\mu=0$ and $\mu_5=0$. Let us denote by $(M_0,\Delta_0)$ the
global minimum point (GMP) of the TDP (\ref{35}). Then, investigating the
behavior of this point vs $\mu$ and $\mu_5$ it is possible to construct the
$(\mu,\mu_5)$-phase portrait (diagram) of the model. A numerical algorithm for
finding the quasi(anti)particle energies  $p_{01}$, $p_{02}$, $\bar p_{01}$,
and $\bar p_{02}$ is elaborated in Appendix B. Based on this, it can be shown
numerically that GMP of the TDP can never be of the form $(M_0\ne 0,\Delta_0\ne
0)$. Hence, at arbitrary fixed values of $M_1$ and $M_2$, i.e. at arbitrary
values of $\delta$ (\ref{20}), it is enough to study the projections
$F_1(M)\equiv\Omega^{ren} (M,\Delta=0)$ and
$F_2(\Delta)\equiv\Omega^{ren}(M=0,\Delta)$ of the TDP (\ref{35}) to the $M$
and $\Delta$ axes, correspondingly. Taking into account the relations
(\ref{36}) and (\ref{37}) for the sum $|p_{01}|+|p_{02}|+|\bar p_{01}|+|\bar
p_{02}|$ at $\Delta=0$ or $M=0$, it is possible  to obtain the following
expressions for these quantities,
\begin{eqnarray}
F_1(M)&=&-\frac{\mu_5^2}{2\pi}-\frac{M^2}{4\pi}+\frac{M^2}{2\pi}\ln\left
(\frac{M}{M_1}\right )-\frac{\theta (\mu-M)}{2\pi}\left
(\mu\sqrt{\mu^2-M^2}-M^2\ln\frac{\mu+\sqrt{\mu^2-M^2}}{M}\right ),
\label{350}\\
F_2(\Delta)&=&-\frac{\mu^2}{2\pi}-\frac{\Delta^2}{4\pi}+\frac{\Delta^2}{2\pi}\ln\left
(\frac{\Delta}{M_2}\right )-\frac{\theta (\mu_5-\Delta)}{2\pi}\left
(\mu_5\sqrt{\mu_5^2-\Delta^2}-\Delta^2\ln\frac{\mu_5+\sqrt{\mu_5^2-\Delta^2}}{\Delta}\right
). \label{035}
\end{eqnarray}
(Details of the derivation of these expressions are given in Appendix
\ref{ApD}.) Now, to find the GMP of the whole TDP (\ref{35}) and, as a
consequence, to obtain the phase structure of the model, it is sufficient to
compare the minimal values of the functions (\ref{350}) and (\ref{035}).
Recall that, up to an unessential constant, each of the functions $F_1(M)$ and
$F_2(\Delta)$ is just a well-known TDP of the usual massless Gross-Neveu model
at zero temperature and nonzero chemical potential. It was investigated, e.g.,
in \cite{Klimenko:1986uq}. So, one can conclude that at $\mu<\mu_c\equiv
M_1/\sqrt{2}$ ($\mu_5<\mu_{5c}\equiv M_2/\sqrt{2}$) the GMP of the function
$F_1(M)$ (of the function $F_2(\Delta$)) lies at the point $M=M_1$ (at the
point $\Delta=M_2$). Whereas at $\mu>\mu_c$ (at $\mu_5>\mu_{5c}$) the GMP is at
the point $M=0$ ($\Delta=0$). Moreover, the corresponding minimal values are
the following:
\begin{eqnarray}
F_1(M_1)=-\frac{\mu_5^2}{2\pi}-\frac{M_1^2}{4\pi},~~
F_2(M_2)=-\frac{\mu^2}{2\pi}-\frac{M_2^2}{4\pi},~~F_1(0)=F_2(0)=-\frac{\mu_5^2}{2\pi}-\frac{\mu^2}{2\pi}.
\label{0035}
\end{eqnarray}
Comparing the least values (\ref{0035}) of the TDPs (\ref{350}) and (\ref{035})
for different values of the chemical potentials $\mu$ and $\mu_5$, it is
possible to obtain the $(\mu,\mu_5)$-phase portrait of the model, which
consists of only three phases, the chiral symmetry breaking phase, the
superconducting phase and, finally, symmetrical phase. Moreover, it is
evident that in the CSB phase the GMP of the TDP (\ref{35}) has the form
$(M_1,0)$, and in the SC phase it lies at the point $(0,M_2)$, whereas in the
symmetrical phase the least value of the TDP (\ref{35}) is reached at the point
$(M=0,\Delta=0)$. Note that the phase structure of the model depends essentially
on the relation between $M_1$ and $M_2$. Indeed, let us first suppose that
$M_1>M_2$. In this case the typical $(\mu,\mu_5)$-phase portrait of the model
is presented in Fig. 1. It is evident that the region $\{\mu>\mu_c,
\mu_5>\mu_{5c}\}$ of the figure corresponds to the symmetrical phase of the
model. Moreover, in the region $\{\mu<\mu_c,\mu_5>\mu_{5c}\}$ (in the region
$\{\mu>\mu_c, \mu_5<\mu_{5c}\}$) of the figure the CSB phase (the SC phase) is
arranged. The competition between CSB and SC phases takes place in the region
$\{\mu<\mu_c, \mu_5<\mu_{5c}\}$. Namely, the critical curve $l$ of Fig. 1 is
defined by the equation $F_1(M_1)=F_2(M_2)$, i.e. by the equation
\begin{eqnarray}
\Omega^{ren} (M=M_1,\Delta=0)&=&\Omega^{ren}(M=0,\Delta=M_2). \label{0350}
\end{eqnarray}
The curve $l$ divides this region into two subregions. To the left of $l$ the
CSB phase is arranged, whereas to the right of $l$ we have the SC phase.
Furthermore, it is clear from (\ref{0350}) and (\ref{0035}) that it is possible
to obtain an exact analytical expression for $l$,
\begin{eqnarray}
l=\left\{(\mu,\mu_5):\mu=\sqrt{\mu_5^2+\frac{M_1^2-M_2^2}{2}}\right\}.
\label{03500}
\end{eqnarray}

In a similar way it is possible to construct a $(\mu,\mu_5)$-phase portrait of
the model when $M_1<M_2$ (the typical $(\mu,\mu_5)$-phase portrait is presented
in Fig. 2). The critical curve $\tilde l$ of the figure is given by the relation
\begin{eqnarray}
\tilde l=\left\{(\mu,\mu_5):\mu_5=\sqrt{\mu^2+\frac{M_2^2-M_1^2}{2}}\right\}.
\label{00350}
\end{eqnarray}
Finally, if $M_1=M_2$, then the typical $(\mu,\mu_5)$-phase portrait of the
model is given in Fig. 3.

Suppose that the values of $M_1$ and $M_2$, for which the phase portrait of
Fig. 2 is drawn, are obtained by rearrangement of the corresponding
$M_1$, $M_2$ values for which Fig. 1 is depicted (and vice versa). For example,
let us assume that $M_1=m_1$, $M_2=m_2$ ($m_1>m_2$) in Fig. 1, but Fig. 2 is
obtained for values $M_1=m_2$ and $M_2=m_1$. Then it is easy to show that Figs.
1 and 2 are dually connected; i.e. Fig. 2 can be obtained from Fig. 1 by
applying the duality transformation $D$ (\ref{28}) (and vice versa). Indeed,
the transformation $D$ can be divided into three more simple steps. i) First,
performing the $\mu\leftrightarrow\mu_5$ transformation in Fig. 1, we rename
the coordinate axes of the figure. ii) Second, when the coordinates of the GMP
are transposed, i.e. $M_0\leftrightarrow \Delta_0$, we have renaming of the
phases. (For example, in this case the GMP of the CSB phase, i.e. the point
$(M_1,0)$, is transformed into the point $(0,M_1)$ and, as a result, the CSB
phase is transformed into the SC phase.) iii) Finally, performing the
transposition $M_1\leftrightarrow M_2$ (which corresponds to
$G_1\leftrightarrow G_2$ of (\ref{28})) and directing vertically (horizontally)
the $\mu_5$ axis (the $\mu$ axis), we obtain just Fig. 2, corresponding to
$M_2=m_1>M_1=m_2$.

It is interesting to note that at $M_1=M_2$, i.e.  at $G_1=G_2$ or $\delta=0$,
the phase portrait of the model (see Fig. 3) is dually invariant, or self-dual.
Moreover, in spite of  self-duality at $G_1=G_2$ of the phase structure of the
model, the CSB and SC ground states are not degenerate in this case. Indeed, at
$\mu_5>\mu$ the CSB phase is preferable, but at $\mu_5<\mu$ the ground state of
the SC phase has a lower energy (at $\mu<\mu_c$). The degeneracy between ground
states of these phases occurs in this case only at the critical curve $L$ (see
Fig. 3), where $\mu=\mu_5$.
\begin{figure}
\includegraphics[width=0.45\textwidth]{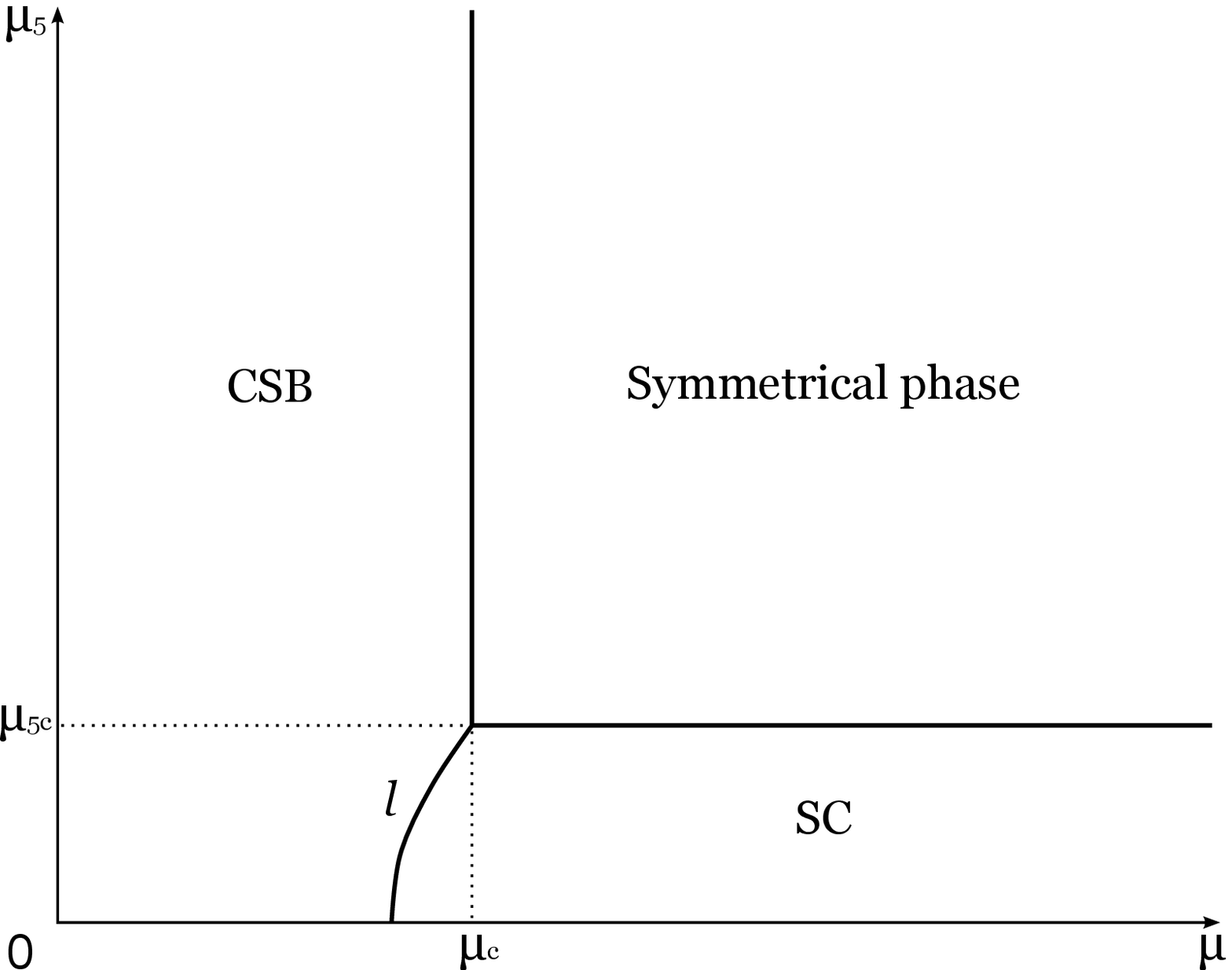}
\hfill
\includegraphics[width=0.45\textwidth]{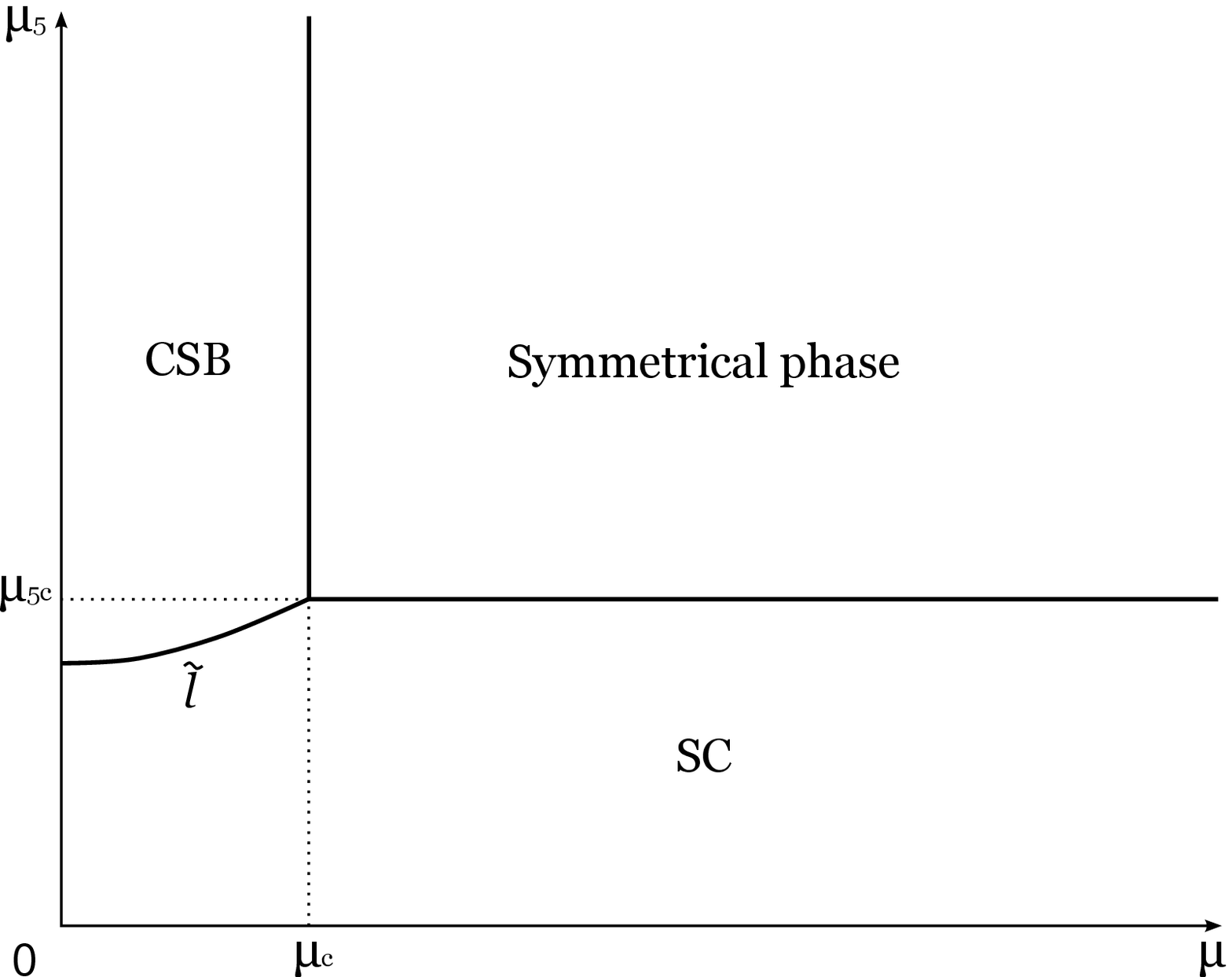}
\\
\parbox[t]{0.45\textwidth}{
\caption{The typical $(\mu,\mu_5)$-phase structure of the model in the
homogeneous case of the ansatz (\ref{8}) for condensates ($b= 0$, $b'= 0$),
when $M_1>M_2$. The notations CSB and SC are used for the chiral symmetry
breaking and superconducting phases, respectively. $\mu_c=M_1/\sqrt{2}$,
$\mu_{5c}=M_2/\sqrt{2}$. The boundary $l$ between CSB and SC phases is defined
by (\ref{03500}).
 } } \hfill
\parbox[t]{0.45\textwidth}{
\caption{The typical $(\mu,\mu_5)$-phase structure of the model in the
homogeneous case of the ansatz (\ref{8}) for condensates ($b= 0$, $b'= 0$),
when $M_1<M_2$. The notations CSB and SC are used for the chiral symmetry
breaking and superconducting phases, respectively. $\mu_c=M_1/\sqrt{2}$,
$\mu_{5c}=M_2/\sqrt{2}$. The boundary $\tilde l$ between CSB and SC phases is
defined by (\ref{00350}). }}
\end{figure}
\begin{figure}
\includegraphics[width=0.45\textwidth]{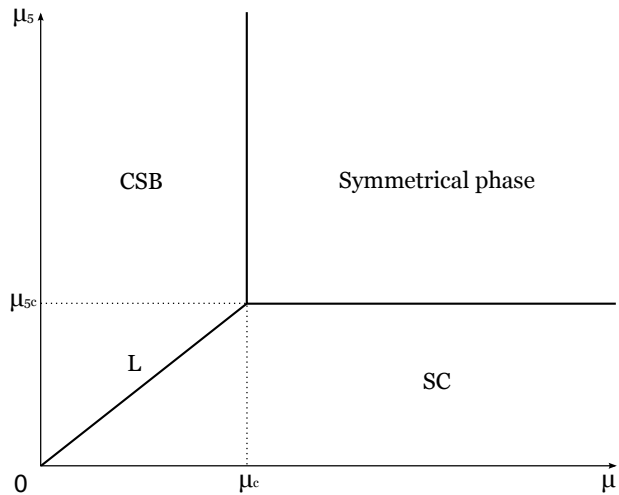}
\\
 \caption{The typical $(\mu,\mu_5)$-phase structure of the model in the
homogeneous case of the ansatz (\ref{8}) for condensates ($b= 0$, $b'= 0$),
when $M_1=M_2$. All the notations are the same as in Figs 1, 2 except the
boundary L between CSB and SC phases, which is defined by the relation
$\mu=\mu_5$.
 } 
\end{figure}

The knowledge of the GMP $(M_0,\Delta_0)$  of the TDP (\ref{35}) provides us
with particle number density $n$ and chiral charge density $n_5$:
\begin{eqnarray}
n=-\frac{\partial\Omega^{ren} (M_0,\Delta_0)}{\partial\mu},~~~~~~
n_5=-\frac{\partial\Omega^{ren} (M_0,\Delta_0)}{\partial \mu_5}. \label{21}
\end{eqnarray}
So, to obtain the behavior of these quantities in the symmetrical, CSB and SC
phases of Figs 1, 2 one can use directly the least values (\ref{0035}) of the
TDP (\ref{035}) in these phases. For example, we have for densities $n$ and
$n_5$ in the CSB phase:
\begin{eqnarray}
n\big |_{\rm CSB}=-\frac{\partial F_1(M_1)}{\partial\mu}\equiv 0,~ n_5\big
|_{\rm CSB}=-\frac{\partial F_1(M_1)}{\partial \mu_5}=\frac{\mu_5}{\pi}.
\label{410}
\end{eqnarray}
By analogy, in the SC and symmetrical (SYM) phases we have for  densities $n$ and $n_5$
\begin{eqnarray}
 n_5\big |_{\rm
SC}\equiv 0,~ n\big |_{\rm SC}=\frac{\mu}{\pi};
~n\big |_{\rm SYM}=\frac{\mu}{\pi},~n_5\big |_{\rm SYM}=\frac{\mu_5}{\pi}.
\label{0410}
\end{eqnarray}
It is clear from (\ref{410}) that at $M_1>M_2$ (or at $G_1>G_2$) the CSB phase
is realized at sufficiently small values of $\mu$ for arbitrary values of
$\mu_5$ (see Fig. 1). Hence, in this case and under a supposition of a
spatially homogeneous structure of the condensates the particle density $n$ of
the system is always equal to zero at sufficiently low values of chemical
potential $\mu$. Correspondingly,  in the case $M_1<M_2$ (or at $G_1<G_2$) the
chiral charge density $n_5$ is equal to zero at sufficiently small values of
$\mu_5$ and for arbitrary values of $\mu$ (see Fig. 2 and (\ref{0410})).

On the basis of the obtained results, we study in the next section the phase
structure of the model when condensates are allowed to be inhomogeneous in the
framework of the ansatz (\ref{8}).

\section{Inhomogeneous case of the ansatz (\ref{8}): $b\ne 0$ and $b'\ne 0$}

 \subsection{Renormalization procedure}

Taking into account the results of the Appendix \ref{ApB}, it is evident that
in the case under consideration the unrenormalized TDP (\ref{11}) can be
obtained from the TDP (\ref{32}), corresponding to the case $b= 0$ and $b'= 0$,
by simple replacements, $\mu\to\tilde\mu\equiv\mu-b$ and
$\mu_5\to\tilde\mu_5\equiv\mu_5-b'$. So we have
\begin{eqnarray}
\Omega^{un}
(M,b,b',\Delta)&=&\frac{M^2}{4G_1}+\frac{\Delta^2}{4G_2}-\Omega_1-\Omega_2-\Omega_3-\Omega_4,
\label{45}
\end{eqnarray}
where
\begin{eqnarray}
\Omega_1&=&\int^\infty_{0}\frac{dp_1}{4\pi}|P_{01}|,~\Omega_2=\int^\infty_{0}\frac{dp_1}{4\pi}|P_{02}|,~
\Omega_3=\int^\infty_{0}\frac{dp_1}{4\pi}|\bar
P_{01}|,~\Omega_4=\int^\infty_{0}\frac{dp_1}{4\pi}|\bar P_{02}|, \label{450}
\end{eqnarray}
and the quantities $P_{01}$, $P_{02}$, $\bar P_{01}$, $\bar P_{02}$ are now the
quasiparticle energies (\ref{B41}), in which the above-mentioned changes of
the chemical potentials should be done, $\mu\to\tilde\mu$ and
$\mu_5\to\tilde\mu_5$. As it follows from the discussion below (\ref{27}), it
is sufficient to study the TDP (\ref{45}) at $M\ge 0$, $\Delta\ge 0$,
$\tilde\mu\ge 0$, and $\tilde\mu_5\ge 0$. Moreover, the TDP (\ref{45}) is
invariant with respect to the duality transformation $\tilde D$:
\begin{eqnarray}
\tilde D:~~G_1\longleftrightarrow G_2,~~\mu\longleftrightarrow\mu_5,~~
M\longleftrightarrow\Delta,~~b\longleftrightarrow b'. \label{280}
\end{eqnarray}
(Recall that after the renormalization procedure the transposition
$G_1\leftrightarrow G_2$ is equivalent to $M_1\leftrightarrow M_2$, or to
changing the sign of the parameter $\delta$ (\ref{20}), $\delta\to -\delta$.)

To find a finite renormalized expression for the TDP (\ref{45}), we should
first regularize it and then perform a renormalization procedure in order to
remove at $\Lambda\to\infty$ the UV divergences by demanding an appropriate
behavior of the bare coupling constants $G_{1,2}$ vs the cutoff parameter
$\Lambda$. In the case of spatially homogeneous condensates all regularization
schemes are usually equivalent. However, in the case of spatially inhomogeneous
condensates the translational invariance over one or several spatial
coordinates is lost. So, the corresponding (spatial) momenta are not conserved.
Then, if one uses the momentum-cutoff regularization technique, as in the
previous section, nonphysical (spurious) $b,b'$-dependent terms appear, and the
TDP acquires some nonphysical properties such as unboundedness from below with
respect to $b$, $b'$, etc. In order to obtain a physically reliable TDP (or
effective potential), in this case an additional substraction procedure is
usually applied (for details see \cite{miransky,gubina}). On the other hand, if
one uses more adequate regularization schemes such as Schwinger proper-time
\cite{nakano,osipov,nickel} or energy-cutoff regularizations
\cite{gubina2,zfk}, etc., such spurious terms do not appear. \footnote{As discussed in the recent papers \cite{gubina,nakano,nickel,gubina2,zfk}, an
adequate regularization scheme in the case of spatially inhomogeneous phases
consists in the following: for different quasiparticles the same restriction
on their region of energy values $|p_{01}|,...,|\bar p_{02}|$ should be used in
a regularized thermodynamic potential.}

In the present paper the energy cutoff regularization scheme of \cite{zfk} is
adopted. (See also \cite{gubina,gubina2,ohwa,Ebert:2013dda}, where a similar
regularization was used in searching for chiral density waves and inhomogeneous
charged pion and Cooper condensates in some NJL$_2$ models.) Namely, we
require that only quasiparticle energies with momenta $p_1$, constrained by
the relations
\begin{eqnarray}
|P_{01}|<\Lambda,~|P_{02}|<\Lambda,~|\bar P_{01}|<\Lambda,~|\bar
P_{02}|<\Lambda,\label{045}
\end{eqnarray}
contribute to the regularized expressions of the integrals (\ref{450}) in
$\Omega_1$,...,$\Omega_4$, correspondingly. At sufficiently high values of the
cutoff $\Lambda$ it is possible to use in (\ref{045}) only the leading terms of
the asymptotic relations (\ref{B26}) for $|P_{01}|$, $P_{02}|$, $|\bar
P_{01}|$, and $|\bar P_{02}|$. As a result, we have the following expressions,
regularized in the framework of the energy cutoff scheme:
\begin{eqnarray}
\Omega^{reg}_1&=&\int^{\Lambda+\tilde\mu-\tilde\mu_5}_{0}\frac{dp_1}{4\pi}|P_{01}|=\int^{\Lambda}_{0}\frac{dp_1}{4\pi}|P_{01}|+\int^{\Lambda+\tilde\mu-\tilde\mu_5}_{\Lambda}\frac{dp_1}{4\pi}|P_{01}|,\nonumber\\
\Omega^{reg}_2&=&\int^{\Lambda-\tilde\mu+\tilde\mu_5}_{0}\frac{dp_1}{4\pi}|P_{02}|=\int^{\Lambda}_{0}\frac{dp_1}{4\pi}|P_{02}|+\int^{\Lambda-\tilde\mu+\tilde\mu_5}_{\Lambda}\frac{dp_1}{4\pi}|P_{02}|,\nonumber\\
\Omega^{reg}_3&=&\int^{\Lambda-\tilde\mu-\tilde\mu_5}_{0}\frac{dp_1}{4\pi}|\bar
P_{01}|=\int^{\Lambda}_{0}\frac{dp_1}{4\pi}|\bar
P_{01}|+\int^{\Lambda-\tilde\mu-\tilde\mu_5}_{\Lambda}\frac{dp_1}{4\pi}|\bar P_{01}|,\nonumber\\
\Omega^{reg}_4&=&\int^{\Lambda+\tilde\mu+\tilde\mu_5}_{0}\frac{dp_1}{4\pi}|\bar
P_{02}|=\int^{\Lambda}_{0}\frac{dp_1}{4\pi}|\bar
P_{02}|+\int^{\Lambda+\tilde\mu+\tilde\mu_5}_{\Lambda}\frac{dp_1}{4\pi}|\bar
P_{02}|. \label{0450}
\end{eqnarray}
Using these expressions instead of $\Omega_i$ in (\ref{45}) ($i=1,...,4$), one
can obtain the following regularized TDP,
\begin{eqnarray}
\Omega^{reg}(M,b,b',\Delta)=\widetilde\Omega^{reg}(M,\Delta)&-&
\int^{\Lambda+\tilde\mu-\tilde\mu_5}_{\Lambda}\frac{dp_1}{4\pi}|P_{01}|-\int^{\Lambda-\tilde\mu+\tilde\mu_5}_{\Lambda}\frac{dp_1}{4\pi}|P_{02}|\nonumber\\&-&
\int^{\Lambda-\tilde\mu-\tilde\mu_5}_{\Lambda}\frac{dp_1}{4\pi}|\bar
P_{01}|-\int^{\Lambda+\tilde\mu+\tilde\mu_5}_{\Lambda}\frac{dp_1}{4\pi}|\bar P_{02}|,
\label{023}
\end{eqnarray}
where $\widetilde\Omega^{reg}(M,\Delta)$ is the TDP (\ref{230}) of the previous
section, regularized by a momentum cutoff approach, in which the replacements
$\mu\to\tilde\mu$ and $\mu_5\to\tilde\mu_5$ should be performed. It is evident
that in the limit $\Lambda\to\infty$ we obtain  from
$\widetilde\Omega^{reg}(M,\Delta)$ the renormalized TDP
$\widetilde\Omega^{ren}(M,\Delta)$ which is the  TDP (\ref{35}), obtained for
the case of homogeneous condensates with $\mu\to\tilde\mu$ and
$\mu_5\to\tilde\mu_5$. So, in the limit $\Lambda\to\infty$ we get from
(\ref{023}) the following expression for the renormalized TDP in the case of
inhomogeneous condensates,
\begin{eqnarray}
\Omega^{ren}(M,b,b',\Delta)=\widetilde\Omega^{ren}(M,\Delta)+\frac{\tilde\mu^2}{2\pi}+\frac{\tilde\mu_5^2}{2\pi}-\frac{\mu^2}{2\pi}-\frac{\mu_5^2}{2\pi}.
\label{46}
\end{eqnarray}
(To obtain the second and third terms in the right-hand side of (\ref{46}), one
should take into account that at $\Lambda\to\infty$ it is possible to use in
(\ref{023}) the asymptotic expansions (\ref{B26}) for the integrand functions
$P_{01}$, $P_{02}$, $\bar P_{01}$, $\bar P_{02}$. Then the integration can be easily
done.  Moreover, we also add to the expression (\ref{46}) unessential
$b,b'$-independent terms, -$\mu^2/2\pi$ and  -$\mu_5^2/2\pi$, in order to
reproduce at $b,b'=0$ the TDP (\ref{35}), corresponding to a spatially
homogeneous chiral condensate. 
\footnote{In fact these summands may be obtained as a result of subtracting the terms
with $b=0$ and $b'=0$ in evaluating Eqs.(\ref{B9}) with the help of a
symbolic formula (\ref{B15}), which  is itself defined up to an
appropriate subtraction (see, e.g. \cite{VCHZH}, p. 248). On the other hand, it is obvious that such a ``by-hand'' addition of last two
terms in (\ref{46}) does not influence the phase structure of
the model. However, we guess that this ``by-hand`` subtracting
procedure could be avoided in the framework of our approach only in
the case of taking into account the factor arising in the path
integral (\ref{9})--(\ref{10}) after the Weinberg transformation (see
the comments in the footnote \ref{fujik}). This factor could have an
appropriate $\mu$ and $\mu_5$ dependence in order to reproduce the
correct expression (\ref{46}) for the TDP.}) \label{4A} 

\subsection{Phase structure at $T=0$}

It is clear that to find the phase portrait of the model at $T=0$,  one should
investigate the global minimum point (GMP) of the TDP
$\Omega^{ren}(M,b,b',\Delta)$ (\ref{46}) vs the dynamical variables
$M,b,b',\Delta$. Since in our case the variables $b$ and $b'$ are absorbed by
the chemical potentials $\mu$ and $\mu_5$, the TDP (\ref{46}) is indeed a
function of four variables $M$, $\Delta$, $\tilde\mu\equiv\mu-b$ and
$\tilde\mu_5\equiv\mu_5-b'$. Thus,  searching for the GMP of this function
consists effectively of two stages. First, one can find the extremum of this
function over $M\ge 0$ and $\Delta\ge 0$ (taking into account the results of
Sec. \ref{mu}) \footnote{As in the case with $b=0$ and $b'=0$, in the
inhomogeneous case we could not find local minimum points of the TDP
(\ref{46}), in which both $M\ne 0$ and $\Delta\ne 0$. } and then one should
minimize the obtained expression over the variables $\tilde\mu\ge 0$,
$\tilde\mu_5\ge 0$. Following this strategy, let us introduce for arbitrary
fixed values of the usual chemical potentials $\mu$ and $\mu_5$ the quantity
\begin{eqnarray}
\omega(\tilde\mu,\tilde\mu_5)=\operatornamewithlimits{min}_{M\ge 0,\Delta\ge
0}\Big\{\Omega^{ren} (M,b,b',\Delta)\Big\}. \label{40}
\end{eqnarray}
Taking into account the results of the investigation of the GMP of the TDP
(\ref{35}) (see Sec. \ref{mu}), it is easy to show that if a point
$(\tilde\mu,\tilde\mu_5)$ belongs to the CSB regions of the
$(\tilde\mu,\tilde\mu_5)$ plane (see, e.g, Fig. 1 with replacements
$\mu\to\tilde\mu$ and $\mu_5\to\tilde\mu_5$), then, as it follows from
(\ref{0035}), we have for the function (\ref{40})
\begin{eqnarray}
\omega(\tilde\mu,\tilde\mu_5)\big |_{\rm
CSB}&=&\frac{\tilde\mu^2}{2\pi}-\frac{M_1^2}{4\pi}-\frac{\mu^2}{2\pi}-\frac{\mu_5^2}{2\pi}.
\label{3500}
\end{eqnarray}
In a similar way, it is easily seen from (\ref{40}), (\ref{46}) and
(\ref{0035}) that if a point $(\tilde\mu,\tilde\mu_5)$ lies in the SC or
symmetrical region of the above-mentioned
$(\tilde\mu,\tilde\mu_5)$ plane, then the function
$\omega(\tilde\mu,\tilde\mu_5)$ is reduced to the expressions
\begin{eqnarray}
\omega(\tilde\mu,\tilde\mu_5)\big |_{\rm
SC}&=&\frac{\tilde\mu_5^2}{2\pi}-\frac{M_2^2}{4\pi}-\frac{\mu^2}{2\pi}-\frac{\mu_5^2}{2\pi},~
\omega(\tilde\mu,\tilde\mu_5)\big |_{\rm SYM}=
-\frac{\mu^2}{2\pi}-\frac{\mu_5^2}{2\pi},\label{1350}
\end{eqnarray}
correspondingly. (Note that in (\ref{3500}) and (\ref{1350}) the  chemical
potentials without tildes, $\mu$ and $\mu_5$, are some fixed external parameters.) It is
evident that the function $\omega(\tilde\mu,\tilde\mu_5)$, presented by the
expressions (\ref{3500}) and (\ref{1350}), is a continuous one in the region
$\tilde\mu\ge 0,\tilde\mu_5\ge 0$. Further, to find the least value of this
function over variables $\tilde\mu\ge 0$ and $\tilde\mu_5\ge 0$ as well as the
points where it is achieved, we consider three qualitatively different cases,
(i) $M_1>M_2$, (ii) $M_1<M_2$, and (iii) $M_1=M_2$.

{\bf (i) The case $M_1>M_2$ ($G_1>G_2$).} In this case it is easy to see from
the relations (\ref{3500})-(\ref{1350}) that the function (\ref{40})
$\omega(\tilde\mu,\tilde\mu_5)$ reaches its minimal value on the
$\tilde\mu_5$ axis, i.e. at $\tilde\mu\equiv\mu-b=0$. The set of these points
lies in the CSB region of the $(\tilde\mu,\tilde\mu_5)$ plane corresponding to
the $(M_0=M_1,\Delta_0=0)$-extreme point of the TDP (\ref{35}). Since in this
case the modulus of the difermion condensate is equal to zero, $\Delta_0=0$, we
are free to put $b'=0$, i.e. $\tilde\mu_5=\mu_5$. Hence, at $M_1>M_2$ and at
arbitrary fixed values of chemical potentials $\mu\ge 0$, $\mu_5\ge 0$ the global
minimum of the TDP (\ref{46}) $\Omega^{ren} (M,b,b',\Delta)$ is arranged at the
point $(M=M_1,b=\mu,b'=0,\Delta=0)$, where
\begin{eqnarray}
\Omega^{ren}
(M=M_1,b=\mu,b'=0,\Delta=0)=\omega(\tilde\mu=0,\tilde\mu_5=\mu_5)=-\frac{M_1^2}{4\pi}-\frac{\mu^2}{2\pi}-\frac{\mu_5^2}{2\pi}.
\label{4100}
\end{eqnarray}
As a result, one can see that for arbitrary values of $\mu\ge 0$, $\mu_5\ge 0$
the spatially inhomogeneous phase in the form of chiral spirals (chiral density
waves) is more preferable in the model than any of the three homogeneous phases
(symmetrical, homogeneous chiral symmetry breaking and homogeneous
superconducting phases) or inhomogeneous superconducting phase.

Taking into account the definitions of the particle number density $n$ and
chiral charge density $n_5$ (\ref{21}), it is possible, using the least value
(\ref{4100}) of the TDP (\ref{46}), to find these quantities in the
inhomogeneous chiral density wave phase. Namely, we have in this phase
\begin{eqnarray}
n=\frac{\mu}{\pi},~~~~~~n_5=\frac{\mu_5}{\pi}. \label{47}
\end{eqnarray}
Let us compare the relations (\ref{47}) with expressions
(\ref{410})-(\ref{0410}), obtained for $n$ and $n_5$ densities in the case of homogeneous condensates. We see that at $M_1>M_2$ and in the supposition of
spatially homogeneous condensates the particle density $n$ of the system always
vanishes in the CSB phase, i.e. at sufficiently small values of $\mu$ (see Fig.
1 and (\ref{410})). In contrast, if spatial inhomogeneity of condensates is
allowed in the framework of the model (1) at $\delta>0$, then in its ground
state, corresponding to a chiral density wave phase (at arbitrary values of
$\mu>0$ and $\mu_5>0$), a nonzero particle density $n$ is generated in the
system even at infinitesimal values of $\mu$, as it follows from (\ref{47}).

{\bf (ii) The case $M_1<M_2$ ($G_1<G_2$).} There is no need to study the phase
structure of the model in this case as detailed as at $M_1>M_2$, because the
phase structure of the model at $G_1<G_2$ can be obtained using the invariance
of the TDP (\ref{46}) with respect to the duality transformation (\ref{280}).
Indeed, at $M_1>M_2$, i.e. at $G_1>G_2$, the TDP (\ref{46}) has a global
minimum at the point of the form $(M=M_1,b=\mu,b'=0,\Delta=0)$. Applying the
duality transformation (\ref{280}) to this TDP, i.e. performing the
replacements $G_1\leftrightarrow G_2$ or $M_1\leftrightarrow M_2$,
$b\leftrightarrow b'$, etc., we obtain the TDP, corresponding to the case
$M_1<M_2$, whose least value is achieved at the point
$(M=0,b=0,b'=\mu_5,\Delta=M_2)$, corresponding to a ground state of the
inhomogeneous superconducting phase. Hence, at $M_1<M_2$ for arbitrary values
of $\mu>0$ and $\mu_5>0$ the nonuniform SC phase is realized in the model. The
expressions for particle density $n$ and chiral charge density $n_5$ in this
phase are still represented by the relations (\ref{47}).

Recall that in the case of a homogeneous ansatz for condensates and at $M_1<M_2$ the
superconducting phase with $n_5=0$ is arranged at rather small values of an
axial chemical potential $\mu_5$ (see Fig. 2). However, if the possibility of
spatial inhomogeneous condensates in the form (\ref{8}) is taken into account,
then at $M_1<M_2$ the nonuniform superconducting phase is realized, in which
$n_5\ne 0$ even at arbitrary low values of $\mu_5$.

{\bf (iii) The case $M_1=M_2$ ($G_1=G_2$).} In this case, using the technique of
point {\bf i)}, it is possible to show that at arbitrary fixed $\mu>0$ and
$\mu_5>0$ the TDP (\ref{46}), $\Omega^{ren} (M,b,b',\Delta)$, has a degenerated
least value, which is reached in two different points,
$(M=M_1,b=\mu,b'=0,\Delta=0)$ and $(M=0,b=0,b'=\mu_5,\Delta=M_1)$,
corresponding to ground state expectation values of inhomogeneous chiral
symmetry breaking and superconducting phases. It means that at $M_1=M_2$ there
is a degeneracy between inhomogeneous chiral symmetry breaking and
inhomogeneous superconductivity in the whole $(\mu,\mu_5)$ plane. In contrast,
in the homogeneous case of the ansatz (\ref{8}) for condensates a degeneracy
between spatially uniform CSB and SC phases is absent, except the line
$\mu=\mu_5$ of this plane, where $\mu<M_1/\sqrt{2}$.

The degeneracy of these ground states means that for arbitrary fixed values of
chemical potentials $\mu>0$ and $\mu_5>0$ in the space, filled with the chiral
density wave phase, a bubble of the inhomogeneous superconducting phase (and
vice versa) can be created.

\subsection{Phase structure at $T>0$}

To introduce finite temperature into the above consideration, it is very
convenient to use the following representation of the unrenormalized TDP
(\ref{45}):
\begin{eqnarray}
\Omega^{un} (M,b,b',\Delta)&=& \frac{M^2}{4G_1}+\frac{\Delta^2}{4G_2}
+\frac{i}{2}\int\frac{d^2p}{(2\pi)^2}\ln\Big [\left (p_0-P_{01}\right )\left
(p_0-P_{02}\right)\left (p_0-\bar P_{01}\right)\left (p_0-\bar
P_{02}\right)\Big ]. \label{55}
\end{eqnarray}
(Integrating in (\ref{55}) over $p_0$ with the help of relation (\ref{B15}),
one obtains the expression (\ref{45}) for the unrenormalized TDP.) Then, to
find the temperature-dependent unrenormalized TDP
$\Omega^{un}_{\scriptscriptstyle{T}}(M,b,b',\Delta)$ one should replace in
(\ref{55}) the integration over $p_0$ in favor of the summation over Matsubara
frequencies $\omega_n$ by the rule
\begin{eqnarray}
\int_{-\infty}^{\infty}\frac{dp_0}{2\pi}\big (\cdots\big )\to
iT\sum_{n=-\infty}^{\infty}\big (\cdots\big ),~~~~p_{0}\to p_{0n}\equiv
i\omega_n \equiv i\pi T(2n+1),~~~n=0,\pm 1, \pm 2,... . \label{190}
\end{eqnarray}
Summing over Matsubara frequencies in the obtained expression (see e.g.
\cite{jacobs} and Appendix \ref{ApC}), we have
\begin{eqnarray}
\Omega^{un}_{\scriptscriptstyle{T}}(M,b,b',\Delta)\!
&=&\frac{M^2}{4G_1}+\frac{\Delta^2}{4G_2}-\int^\infty_{0}\frac{dp_1}{4\pi}\Big\{|P_{01}|+|P_{02}|+|\bar
P_{01}|+|\bar P_{02}|\Big\}\nonumber\\
&& -T\int_{0}^{\infty}\frac{dp_1}{2\pi}\ln\left\{\big [1+e^{-\beta
|P_{01}|}\big ]\big [1+ e^{-\beta |P_{02}|}\big ]\big [1+e^{-\beta |\bar
P_{01}|}\big ]\big [1+e^{-\beta |\bar P_{02}|}\big ]\right\}, \label{430}
\end{eqnarray}
where $\beta=1/T$. The last integral in (\ref{430}) is a convergent one,
whereas other terms form the zero temperature unrenormalized TDP (\ref{45}).
Hence, it is sufficient to renormalize just this component of the whole TDP
(\ref{430}), using the energy-cutoff regularization scheme of the previous
Sec. \ref{4A}. As a result, one can obtain finite  and renormalized
expression for the TDP at nonzero $T$,
\begin{eqnarray}
\Omega^{ren}_{\scriptscriptstyle{T}}(M,b,b',\Delta)\!
&=&\Omega^{ren}(M,b,b',\Delta)\nonumber\\&-&
T\int_{0}^{\infty}\frac{dp_1}{2\pi}\ln\left\{\big [1+e^{-\beta |P_{01}|}\big
]\big [1+e^{-\beta |P_{02}|}\big ]\big [1+e^{-\beta |\bar P_{01}|}\big ]\big
[1+e^{-\beta |\bar P_{02}|}\big ]\right\}, \label{57}
\end{eqnarray}
where $\Omega^{ren}(M,b,b',\Delta)$ is the zero temperature TDP (\ref{46}).
Based on the numerical algorithm for finding the quasiparticle energies
$P_{01}$, $P_{02}$, $\bar P_{01}$, $\bar P_{02}$ (see Appendix B), it is possible to
show that at fixed values of the variables $\tilde\mu$ and $\tilde\mu_5$ the
least value of the TDP (\ref{57}) can never be achieved at 
the point $(M,\Delta)$ with both nonzero coordinates, $M\ne 0$ and
$\Delta\ne 0$. 
So to investigate the global minimum of this TDP it is
sufficient to deal with the restrictions of the TDP (\ref{57}) on the manifolds
$\Delta =0$ and $M=0$, i.e. with the quantities
\begin{eqnarray}
\Omega_{1\scriptscriptstyle{T}}(M,b,b')\equiv\Omega^{ren}_{\scriptscriptstyle{T}}(M,b,b',\Delta=0),~~~
\Omega_{2\scriptscriptstyle{T}}(\Delta,b,b')\equiv\Omega^{ren}
_{\scriptscriptstyle{T}}(M=0,b,b',\Delta), \label{58}
\end{eqnarray}
correspondingly. Note that at $\Delta=0$ we have from (\ref{26}) that each of
quasiparticle energies $P_{01}$, $P_{02}$, $\bar P_{01}$, and $\bar P_{02}$ is
equal to one of the expressions $\tilde\mu\pm\sqrt{M^2+(p_1-\tilde\mu_5)^2}$ or
$-\tilde\mu\pm\sqrt{M^2+(p_1+\tilde\mu_5)^2}$, whereas at $M=0$ one can easily
see from (\ref{27}) that each of these quantities is represented by one of the
expressions $\tilde\mu_5\pm\sqrt{\Delta^2+(p_1-\tilde\mu)^2}$ or
$-\tilde\mu_5\pm\sqrt{\Delta^2+(p_1+\tilde\mu)^2}$. Then, one should take into
account the expression (\ref{46}) for the TDP $\Omega^{ren}(M,b,b',\Delta)$ as
well as the relations (\ref{350}) and (\ref{035}) for particular values of the
TDP (\ref{35}) at $\Delta=0$ and $M=0$. Finally, when converting the integral
term of (\ref{57}) we use essentially the following relation
\begin{eqnarray*}
\ln\left (1+e^{-x}\right )=-x+\ln\left (1+e^{x}\right ).
\end{eqnarray*}
As a result, we obtain the following expressions for the TDPs (\ref{58}):
\begin{eqnarray}
\Omega_{1\scriptscriptstyle{T}}(M,b,b')&=&\frac{M^2}{2\pi}\ln\left
(\frac{M}{M_1}\right )-\frac{M^2}{4\pi}+\frac{\tilde\mu^2}{2\pi}-
\frac{\mu^2}{2\pi}-\frac{\mu_5^2}{2\pi}\nonumber\\
&&-T\int_{0}^{\infty}\frac{dq}{\pi}\ln\left\{\left [1+e^{-\beta\left
(\sqrt{M^2+q^2}+\tilde\mu\right )}\right ] \left [1+e^{-\beta\left
(\sqrt{M^2+q^2}-\tilde\mu\right )}\right ]\right\},
 \label{440}\\
\Omega_{2\scriptscriptstyle{T}}(\Delta,b,b')&=& \frac{\Delta^2}{2\pi}\ln\left
(\frac{\Delta}{M_2}\right
)-\frac{\Delta^2}{4\pi}+\frac{\tilde\mu_5^2}{2\pi}-\frac{\mu^2}{2\pi}-
\frac{\mu_5^2}{2\pi}\nonumber\\
&&-T\int_{0}^{\infty}\frac{dq}{\pi}\ln\left\{\left [1+e^{-\beta\left
(\sqrt{\Delta^2+q^2}+\tilde\mu_5\right )}\right ]\left [1+e^{-\beta\left
(\sqrt{\Delta^2+q^2}-\tilde\mu_5\right )}\right ]\right\}. \label{044}
\end{eqnarray}
Note that the function (\ref{440}) (the function (\ref{044})) does not depend on
the variable $b'$ (variable $b$). Due to this fact, it is possible to establish
that the TDP (\ref{440}) has two stationary points, $(M=M_0(T),b=\mu,b'=0)$ and
$(M=0,b=0,b'=0)$, where $M_0(T)$ vs $T$ behaves like the gap in ordinary
Gross-Neveu model with zero chemical potential and $T\ne 0$ \cite{jacobs}, i.e.
$M_0(0)=M_1$ and $M_0(T_{c1})=0$, where $T_{c1}=M_1e^\gamma/\pi$ (here $\gamma$
is the Euler's constant, $\gamma=0.577...$). By analogy, the TDP (\ref{044})
also has two stationary points, $(\Delta=\Delta_0(T),b=0,b'=\mu_5)$ and
$(\Delta=0,b=0,b'=0)$, with similar properties of the gap $\Delta_0(T)$ vs $T$:
$\Delta_0(0)=M_2$ and $\Delta_0(T_{c2})=0$, where $T_{c2}=M_2e^\gamma/\pi$.
Comparing the values of the TDPs (\ref{440}) and (\ref{044}) in the above-mentioned stationary points, it is possible to find the genuine GMP of the
initial TDP (\ref{57}) and, as a consequence, to establish the phase structure
of the model at each fixed value of chemical potentials and temperature. It
turns out that at $M_1>M_2$ the inhomogeneous chiral symmetry breaking (or
chiral density wave) phase is realized in the model at $T<T_{c1}$ for arbitrary
$\mu>0$ and $\mu_5>0$ values. However, at $T>T_{c1}$ one can observe in this
case the symmetrical phase. In contrast, at $M_1<M_2$ the dual phase portrait
is realized in the model: in this case we have an inhomogeneous superconducting
phase at $T<T_{c2}$ and a symmetrical phase at $T>T_{c2}$. If $M_1=M_2$, then at
$T<T_{c1}$ there is a degeneracy between inhomogeneous CSB and inhomogeneous SC
phases, whereas at $T>T_{c1}$ the symmetrical phase is realized.

Finally, a few words are in order about the behavior of the particle number $n$ and
chiral charge $n_5$ densities at nonzero temperature. Recall that  to find
these quantities we should first of all obtain the value of the TDP
(\ref{57}) in its global minimum point
$(M_0,b_0,b'_0,\Delta_0)$. Then, particle number density  $n$ (chiral
charge density $n_5$) is the derivative of the quantity
$\Omega^{ren}(M_0,b_0,b'_0,\Delta_0)$ with respect to chemical
potential $\mu$ (chemical potential $\mu_5$). Hence, taking into
account the above consideration of the phase structure, it is possible
to conclude that both at $M_1>M_2$ and $M_1<M_2$ the same simple
expressions for densities, 
\begin{eqnarray}
n=\frac{\mu}{\pi},~~~n_5=\frac{\mu_5}{\pi}, \label{T}
\end{eqnarray}
are valid for arbitrary temperatures, i.e. in the symmetric phase, in
the inhomogeneous chiral symmetry breaking phase (at $T<T_{c1}$ if
$M_1>M_2$) and in the inhomogeneous superconducting phase (at
$T<T_{c2}$ if $M_1<M_2$). 

\section{Summary and discussions}

In this paper, some thermodynamical properties of the (1+1)-dimensional system,
which is characterized by ground states with nonzero particle number as well as
the chiral charge densities, are considered. The microscopic Lagrangian,
describing physics of the system, is chosen in the form (1); i.e. we deal with
the (1+1)-dimensional NJL model, containing two types, or channels, of
interaction. In the first, chiral, channel the interaction between particles
and antiparticles is characterized by coupling constant $G_1$, whereas in the
second, superconducting channel, we have particle-particle interaction with
coupling $G_2$. The phase structure of the model is investigated in the paper
in terms of particle number $\mu$ and chiral charge $\mu_5$ chemical
potentials. 

There are several reasons for taking into consideration two types of chemical potentials. The first and the 
most important one is to bring the
investigation of the duality between chiral symmetry breaking and
superconductivity to a single platform, i.e. to extend the
investigation of the duality to the framework of a more general
(1+1)-dimensional model (1), rather than the way it was done earlier in \cite{thies1}. 
Recall that in \cite{thies1}, a connection was found 
(duality) between properties of two different models, the GN model
with chemical potential $\mu$, describing quark interaction in the
$q\bar q$ channel only, and the GN model with chemical potential
$\mu_5$, describing the interaction in the $q q$ channel only. In
contrast, in our model (1) there is a competition between these two types
of interaction and, in addition, there are both types of chemical
potentials, $\mu$ and $\mu_5$. The second reason is already a
``traditional'' motivation, which is common in all investigations of
low-dimensional theories, i.e. the possibility and the hope to perform
a deeper consideration of a new physical phenomena in terms of toy
models. In our case these are the parity breaking
\cite{andrianov,andrianov2} and the chiral magnetic \cite{ruggieri}
effects of dense quark-gluon matter, accompanied by a nonzero chiral
charge density $n_5$ (or $\mu_5\ne 0$). 

Moreover, the finite  temperature effect is also taken into
account. It is well known that in any dense system there might appear a
spontaneous breaking of spatial translational invariance, resulting in a spatial
dependence of order parameters, or condensates. So we investigated a phase
structure of the model, assuming the Fulde-Ferrel \cite{ff} single plane wave
ansatz (\ref{8}) for condensates. (In particular, for the chiral condensate the
ansatz (\ref{8}) is known as a chiral density wave or chiral spiral.) For
comparison, we investigate a phase structure of the model in two particular
cases of the ansatz (\ref{8}): (i) when $b=0$ and $b'=0$, i.e. the condensates
are put as spatially homogeneous by hand, and (ii) when the parameters $b$, $b'$
are dynamical quantities, defined by gap equations. The main results of the
paper are the following.

(1) First of all, we have established that in the homogeneous case of the ansatz
(\ref{8}) for condensates (at $b=0$ and $b'=0$) the thermodynamic potential
of the model is invariant under the duality transformation $D$
(\ref{28}). It means that if at $G_1>G_2$ (or, equivalently, at $M_1>M_2$,
where the connections between $G_{1,2}$ and $M_{1,2}$ are represented in
(\ref{31})) the CSB phase (SC phase) is realized in the model at some fixed
values of chemical potentials, e.g., at $\mu=\alpha$, $\mu_5=\beta$, then at
$G_1\leftrightarrow G_2$ the system is in the SC phase (CSB phase) at
$\mu=\beta$, $\mu_5=\alpha$. Taking into account this duality correspondence
property of the model, it is sufficient to study the $(\mu,\mu_5)$-phase
diagram only at $G_1>G_2$, i.e. at $M_1>M_2$ (see, e.g., Fig. 1). Then the
phase portrait of the model at $M_1<M_2$ (see Fig. 2) is simply the dual
mapping of Fig. 1.

(2) At $G_1=G_2$ (or at $M_1=M_2$) the $(\mu,\mu_5)$-phase diagram of the model
in the homogeneous case of the ansatz (\ref{8}) for condensates is presented in
Fig. 3. Clearly, this diagram is invariant with respect to  the duality
transformation $D$ (\ref{28}); i.e. one can say that the model is self-dual
in this case. Nevertheless, we would like to emphasize that in the homogeneous
case of the ansatz (\ref{8}) and at  $\mu\ne \mu_5$ the self-duality property
of the model does not mean the degeneracy of the CSB and SC ground states at
$G_1=G_2$. The CSB-SC degeneracy appears only on the line L of Fig. 3, i.e. at
$\mu=\mu_5$.

(3) If a spatially inhomogeneous behavior of condensates is assumed in the form
(\ref{8}), where the parameters $b$ and $b'$ must be found by gap equations,
then the $(\mu,\mu_5)$-phase structure of the model  is considerably
simplified. Indeed, in this case at $G_1>G_2$, i.e. at $M_1>M_2$, (at
$G_1<G_2$) only the inhomogeneous chiral density wave phase (only inhomogeneous
SC phase) is realized in the model for arbitrary values of $\mu$ and $\mu_5$.
The critical temperature, at which the inhomogeneous chiral density wave phase
(the inhomogeneous SC phase) is destroyed and the symmetrical phase appears, is
equal to $T_{c1}=M_1e^\gamma/\pi$ (equal to $T_{c2}=M_2e^\gamma/\pi$). (In
contrast, if $b$ and $b'$ are equal to zero {\it a priori}, i.e. condensates are
assumed to be homogeneous from the very beginning, then, depending on the
relation between $\mu$ and $\mu_5$, spatially uniform CSB and SC phases are
present on a model phase portrait both at $G_1>G_2$ and $G_1<G_2$ (see, e.g.,
Figs 1, 2).) Note also that if $G_1\ne G_2$, then the inhomogeneous chiral
density wave phase is a dual mapping of the inhomogeneous SC phase and vice
versa. Moreover, in this case the degeneracy between the above-mentioned
inhomogeneous phases is absent.

(4) It is interesting to note that at $G_1=G_2$ and for arbitrary fixed values
of the chemical potentials $\mu$ and $\mu_5$ the self-dual and  degenerated
phase portrait of the above-mentioned inhomogeneous phases appears. It means
that for each fixed value of $\mu$ and $\mu_5$ there is an equal
opportunity for the emergence as one or the other inhomogeneous phase in the
system. Moreover, the coexistence of these phases is not excluded.  In
contrast, in the homogeneous case of the ansatz  (\ref{8}) for condensates the
degeneration between  CSB and SC ground states of the model is absent (at
$\mu\ne \mu_5$), in spite of a self-dual phase portrait of the model at
$G_1=G_2$ (see Fig. 3).

(5) Note that if the condesates are homogeneous and $T=0$, then in the CSB phase the particle density $n$ is identically zero, whereas the chiral charge density $n_5$ vanishes in the SC phase (see (\ref{410}) and (\ref{0410})). However, if a
possibility of spatial inhomogeneity for condensates in the form (\ref{8}) is
taken into account, then both in the ground state corresponding to an
inhomogeneous chiral density wave phase and in the inhomogeneous SC phase the
nonzero particle number density, $n=\mu/\pi$, and nonzero chiral charge density, $n_5=\mu_5/\pi$,  appear. 
Moreover, at $T\ne 0$ and in the case of inhomogeneous ansatz
(\ref{8}) for condensates both particle density $n$ and chiral charge
density $n_5$ do not depend on temperature and have the same behaviors
(\ref{T}) in all possible phases of the model, symmetrical,
inhomogeneous chiral density wave, and inhomogeneous superconducting
phases. \vspace{0.2cm} 

We are aware of the fact that some of the above properties (such as the appearance of an inhomogeneous phase at arbitrarily low chemical potentials or a change of the nature of the inhomogeneous phase (CSB vs SC) if the SC coupling becomes larger than the CSB one) are peculiarities of the above-considered (1+1)-dimensional model (1) and, perhaps, have no relation to reality. 
However, there are results (e.g. the extension of inhomogeneous phases to high values of chemical potentials) that are predicted by (3+1)-dimensional QCD-like models as well \cite{buballa}. 

It is also worth noting that in the recent paper \cite{andrianov2} the (3+1)-dimensional NJL model with several quark-antiquark interaction channels was investigated at zero temperature and in the presence of two chemical potentials, $\mu$ and $\mu_5$. The only homogeneous ansatz for condensates is taken into account in this research, which is devoted to the study of the parity breaking effects in QCD at high densities.
As it was established there, in dense quark matter, i.e. at $\mu\ne 0$, and at a rather high values of $\mu_5$ the chiral symmetry breaking+parity breaking phase is allowed to exist (see Fig. 7 in this paper). In some ways our work is related to the same problem, but only considered in the framework of a simple NJL$_2$ toy model (1). Indeed, we have shown that at $T<T_{c1}$ and $G_1>G_2$ an inhomogeneous CSB phase, in which parity is also spontaneously broken down, is realized in the model (1).
So, there is an alternative mechanism to achieve parity breaking in dense QCD, based on spatially nonuniform condensates. \footnote{As it follows from our consideration, the role of the chiral chemical potential $\mu_5$ in this approach is simply to supply a nonzero chiral charge density $n_5=\mu_5/\pi$ (\ref{T}) in the chiral density wave phase, where both parity and chiral symmetry are broken spontaneously. }   
Moreover, we have demonstrated that, in comparison with the NJL model \cite{andrianov2}, a reduced number of quark-antiquark channels of interaction is needed in order to obtain spontaneous parity and chiral symmetry breaking at $\mu\ne 0$. 

Finally, note that inhomogeneous phases are observed in a phase diagram of the NJL$_2$ model (1) only at small temperatures, i.e. at $T<T_{c1}$ if $G_1>G_2$ or at $T<T_{c2}$ if $G_1<G_2$. At high temperatures the symmetric phase with  $n\ne 0$ and, especially, with nonzero chiral charge density $n_5$ is realized (see in Sec. IV C). It is well known that in a heavy ion collision scenario both the temperature and magnetic field can be extremely high. Due to high temperatures a sphaleron transition might occur, which is accompanied by an appearance of a nonzero chiral density $n_5$ (or chemical potential $\mu_5\ne 0$) in the system \cite{ruggieri,huang}. Due to strong magnetic fields the dynamics of the system becomes essentially one dimensional. So we believe that in the high temperature region the model (1) reflects in some details the physics of quark-gluon plasma and, furthermore, the (1+1)-dimensional models with $\mu_5$-chemical potential deserve to be investigated.
\appendix

\section{The path integration over anticommuting fields}
\label{ApA}

Let us calculate the following path integral over anticommuting two-component
Dirac spinor fields $q(x)$, $\bar q(x)$:
\begin{eqnarray}
I=\int[d\bar q][dq]\exp\Big (i\int d^2 x\Big [\bar q  D
q-\frac{\Delta}{2}(q^T\epsilon q) -\frac{\Delta}{2}(\bar q \epsilon\bar
q^T)\Big ] \Big ) \label{A1},
\end{eqnarray}
where we use the notations of Sec. \ref{effaction}. In particular, the
operator $D$ is given in (\ref{110}) and $\epsilon$ is defined in (\ref{2}).
Note that the integral $I$ is equal to the argument of the $\ln (x)$ function in
the formula (\ref{10}) in the particular case $N=1$. Recall that there are general
Gaussian path integrals \cite{vasiliev}:
\begin{eqnarray}
\int[dq]\exp\Big (i\int d^2 x\Big [-\frac{1}{2}q^T A q +\eta^Tq \Big ]\Big
)&=&\left(\det(A)\right )^{1/2}\exp\Big (-\frac{i}{2}\int d^2 x\Big
[\eta^T A^{-1}\eta\Big ]\Big ),  \label{A2}\\
\int[d\bar q]\exp\Big (i\int d^2 x\Big [-\frac{1}{2}\bar q A \bar q^T
+\bar\eta\bar q^T\Big ] \Big )& =&\left (\det(A)\right )^{1/2}\exp\Big
(-\frac{i}{2}\int d^2 x\Big [\bar\eta A^{-1}\bar\eta^T\Big ]\Big ), \label{A3}
\end{eqnarray}
where $A$ is an antisymmetric operator in coordinate and spinor spaces, and
$\eta(x)$, $\bar\eta (x)$ are spinor anticommuting sources which also
anticommute with $q$ and $\bar q$. First, let us integrate in (\ref{A1}) over
$q$ fields with the help of the relation (\ref{A2}) supposing there that
$A=\Delta\epsilon$, $\bar qD=\eta^T$, i.e. $\eta=D^T\bar q^T$. Then
\begin{eqnarray}
I=\left (\det(\Delta\epsilon)\right )^{1/2}\int[d\bar q]\exp\Big
(-\frac{i}{2}\int d^2 x \bar q \big
[\Delta\epsilon+D(\Delta\epsilon)^{-1}D^T\big ]\bar q^T\Big ). \label{A4}
\end{eqnarray}
Second, the integration over $\bar q$ fields in (\ref{A4}) can be easily
performed with the help of the formula (\ref{A3}), where one should put
$A=\Delta\epsilon+D(\Delta\epsilon)^{-1}D^T$ and $\bar\eta=0$. As a result, we
have
\begin{eqnarray}
I=\left (\det(\Delta\epsilon)\right )^{1/2}\left
(\det[\Delta\epsilon+D(\Delta\epsilon)^{-1}D^T]\right )^{1/2}=\left (\det
[-\Delta^2-D\epsilon D^T\epsilon]\right )^{1/2}, \label{A5}
\end{eqnarray}
where we took into account that $\epsilon\epsilon =-1$ and
$\epsilon^{-1}=-\epsilon$. For the following one should remember the well-known
relations: $(\partial_\nu)^T=-\partial_\nu$, $\epsilon (\gamma^\nu)^T\epsilon
=\gamma^\nu$, where $\nu=0,1$. Hence,
\begin{eqnarray}
I=\left (\det [-\Delta^2+D_+D_-]\right )^{1/2}\equiv\left (\det B\right )^{1/2}
, \label{A6}
\end{eqnarray}
where $D_\pm=\gamma^\nu i\partial_\nu-M\pm
((\mu-b)\gamma^0+(\mu_5-b')\gamma^1)$. Using the general relation $\det B =\exp
({\rm Tr}\ln B)$, we get from (\ref{A6}):
\begin{eqnarray}
\ln I=\frac 12 {\rm Tr}\ln\left (-\Delta^2+D_+D_-\right
)=\sum_{i=1}^{2}\int\frac{d^2p}{(2\pi)^2} \ln(\lambda_i(p))\int d^2x.
\label{A7}
\end{eqnarray}
(A more detailed consideration of operator traces is presented in  Appendix
A of the paper \cite{Ebert:2009ty}.) In this formula the symbol Tr means the trace
of an operator both in the coordinate and internal spaces. Moreover,
$\lambda_i(p)$ ($i=1,2$) are eigenvalues of the 2$\times$2 Fourier transformed
matrix $\bar B(p)$ of the operator $B$, i.e.
\begin{eqnarray}
\lambda_{1,2}(p)=p_0^2-\tilde\mu^2-p_1^2+\tilde\mu_5^2+M^2-\Delta^2\pm
2\sqrt{M^2 p_0^2-M^2 p_1^2+\tilde\mu^2 p_1^2-2 p_0\tilde\mu_5\tilde\mu
p_1+p_0^2\tilde\mu_5^2},\label{A8}
\end{eqnarray}
where $\tilde\mu=\mu-b$ and $\tilde\mu_5=\mu_5-b'$.

\section{Evaluation of the TDP (\ref{25})}
\label{ApB}

In order to renormalize and then to investigate the TDP (\ref{25}), it  is
necessary to modify the initial expression (\ref{25}). First let us obtain a
more convenient expression for $\det\bar B(p)$. With this aim we use the
following alternative form of the relation (\ref{26}), namely
\begin{eqnarray}
\det\bar B(p) &=& p_0^4 - 2(M^2+\Delta^2+\mu^2+\mu_5^2+p_1^2)p_0^2+8\mu_5\mu p_1p_0+p_1^4-2p_1^2(\mu^2+\mu_5^2-M^2-\Delta^2)\nonumber\\
&+&(\Delta^2-M^2+\mu^2-\mu_5^2)^2\equiv p_0^4 + \alpha p_0^2 + \beta
p_0+\gamma,\label{B1}
\end{eqnarray}
where, evidently,
\begin{eqnarray}
\alpha &=& -2 (M^2+\Delta^2+\mu^2+\mu_5^2+p_1^2),~~~
\beta= 8\mu_5\mu p_1,\nonumber\\
\gamma &=&
p_1^4-2p_1^2(\mu^2+\mu_5^2-M^2-\Delta^2)+(\Delta^2-M^2+\mu^2-\mu_5^2)^2.\label{B2}
\end{eqnarray}
It is very convenient to present the fourth-order polynomial of the variable
$p_0$ (\ref{B1}) as a product of two second-order polynomials (this way is
proposed in \cite{Birkhoff}); i.e. we assume that
\begin{eqnarray}
p_0^4&+&\alpha p_0^2+\beta p_0 +\gamma = (p_0^2 + r p_0 + q)(p_0^2 - r
p_0 + s)\nonumber\\
&&=\left [\left (p_0+\frac r2\right )^2+q-\frac{r^2}{4}\right ]\left [\left
(p_0-\frac r2\right )^2+s-\frac{r^2}{4}\right ]\equiv
(p_0-p_{01})(p_0-p_{02})(p_0-\bar p_{01})(p_0-\bar p_{02}),\label{B3}
\end{eqnarray}
where $r$, $q$ and $s$ are some real valued quantities, such that
\begin{eqnarray}
\alpha = s+q-r^2,~~ \beta = rs-qr,~~ \gamma= sq.\label{B4}
\end{eqnarray}
Then, using expansion (\ref{B3}), it is easy to present all the roots $p_{01}$,
$p_{02}$, $\bar p_{01}$, and $\bar p_{02}$ of the polynomial
(\ref{B1})-(\ref{B3}) vs $p_0$ in the following form:
\begin{eqnarray}
p_{01}=-\frac{r}{2}+\sqrt{\frac{r^2}{4}-q},~~p_{02}=\frac{r}{2}+
\sqrt{\frac{r^2}{4}-s},~~\bar p_{01}=-\frac{r}{2}-\sqrt{\frac{r^2}{4}-q},~~\bar
p_{02}=\frac{r}{2}-\sqrt{\frac{r^2}{4}-s}.\label{B41}
\end{eqnarray}
The expressions (\ref{B41}) are usually called the dispersion laws (or
relations) of the model. So, the quantities $p_{01}$ and $p_{02}$ are the
energies of  two quasiparticles, whereas $\bar p_{01}$ and $\bar p_{02}$ are
the energies of their quasiantiparticles. Since in (\ref{B3}) the energy
parameter $p_0$ is shifted by $\pm r/2$, one may interpret the quantity $r/2$
as an effective chemical potential. In the following we are going to use just
the quantities (\ref{B41}) in our numerical calculations, so it is necessary to
express the coefficients $r$, $q$ and $s$ in (\ref{B3}) in terms of $\alpha$, $\beta$, $\gamma$-quantities.

Suppose first that $\mu=0$ and $\mu_5=0$ (other variables, $M$, $\Delta$, and
$p_1$, are nonzero). Then, as it is clear from (\ref{29}), $r=0$,
$s=-p_1^2-(M-\Delta)^2$ and $q=-p_1^2-(M+\Delta)^2$. In particular, it means
that in this case
 \begin{eqnarray}
\big (|p_{01}|+|p_{02}|+|\bar p_{01}|+|\bar p_{02}|\big )\big
|_{\mu=0,\mu_5=0}=2\sqrt{p_1^2+(M+\Delta)^2}+2\sqrt{p_1^2+(M-\Delta)^2}.
\label{B42}
\end{eqnarray}
In the general case, when both $\mu\ne 0$ and $\mu_5\ne 0$, one can solve the
system of equations (\ref{B4}) and find
\begin{eqnarray}
q=\frac 12 \left (\alpha +R-\frac{\beta}{\sqrt{R}}\right ),~~ s=\frac 12 \left
(\alpha +R+\frac{\beta}{\sqrt{R}}\right ),~~r=\sqrt{R},\label{B5}
\end{eqnarray}
where $R$ is an arbitrary positive real solution of the equation
\begin{eqnarray}
\label{B6} X^3 + AX=BX^2 + C
\end{eqnarray}
with respect to a variable $X$, and
\begin{eqnarray}
A&=& \alpha^2-4\gamma=16\Big[\mu_5^2\Delta^2+M^2\mu^2+
\Delta^2M^2+\mu_5^2\mu^2+p_1^2(\mu^2+\mu_5^2)\Big],\nonumber\\
B&=&-2\alpha =4(M^2+\Delta^2+\mu^2+\mu_5^2+p_1^2),~~ C=\beta^2=(8\mu_5\mu
p_1)^2. \label{B7}
\end{eqnarray}
Numerical investigation shows that for any fixed values of $\mu>0$, $\mu_5>0$,
$M>0$, $\Delta>0$ and $p_1$ the discriminant of the third-order algebraic
equation (\ref{B6}), i.e. the quantity $18ABC-4B^3C+A^2B^2-4A^3-27C^2$, is
always positive. So the equation (\ref{B6}) vs $X$ has three different real
solutions $R_1,R_2$ and $R_3$ (this fact is presented in \cite{Birkhoff}).
Moreover, since the coefficients $A$, $B$ and $C$ (\ref{B7}) are positively
defined, it is clear that all the roots $R_1$, $R_2$ and $R_3$ are positive
quantities. So we are free to choose the quantity $R$ from (\ref{B5}) as one of
the solutions $R_1$, $R_2$ or $R_3$. In each case, i.e. for $R=R_1$, $R=R_2$, or
$R=R_3$, we will obtain the same set of the roots (\ref{B41}) (possibly
rearranged), which depends only on $\mu$, $\mu_5$, $M$, $\Delta$ and $p_1$, and
does not depend on the choice of $R$.

Using standard methods, it is possible to find the following $p_1\to\infty$
asymptotic expansions of the roots $R_1$, $R_2$ and $R_3$,
\begin{eqnarray}
\label{B701}
R_1&=&4\mu^2+\frac{4\Delta^2\mu^2[\mu^2-M^2-\mu_5^2]}{(\mu_5^2-\mu^2)p_1^2}
+{\cal O}\big (1/p_1^4\big ),\\
\label{B7001}
R_2&=&4\mu_5^2+\frac{4M^2\mu_5^2[\mu_5^2-\Delta^2-\mu^2]}{(\mu^2-\mu_5^2)p_1^2}
+{\cal O}\big (1/p_1^4\big ),\\
R_3&=&4p_1^2+4(M^2+\Delta^2)+\frac{4(\mu_5^2M^2+\mu^2\Delta^2-M^2\Delta^2)}{p_1^2}
+{\cal O}\big (1/p_1^4\big ). \label{B71}
\end{eqnarray}
It is clear from these relations that $R_3$ is invariant, whereas
$R_1\leftrightarrow R_2$ under the duality transformation (\ref{28}). Note that the
expansions (\ref{B701}) and (\ref{B7001}) are valid only at $\mu_5\ne\mu$. If
$\mu_5=\mu$, then at $p_1\to\infty$ we have for $R_{1,2}$ the expansions
\begin{eqnarray}
R_{1,2}=4\mu^2\pm\frac{4\mu\Delta
M}{|p_1|}+\frac{2\Delta^2M^2-2\mu^2\Delta^2-2\mu^2 M^2}{p_1^2}+{\cal O}\big
(1/p_1^3\big ). \label{B071}
\end{eqnarray}
(In this particular case each of the roots $R_{1,2,3}$ is invariant with
respect to the duality transformation (\ref{28}).) It was mentioned above that
the quantity $r/2$ can be interpreted as an effective chemical potential (see
the text after (\ref{B41})). Moreover, it is clear from (\ref{B701}) that just
the choice $R=R_1$ supports this interpretation, since at $p_1\to\infty$ we
have $r/2=\sqrt{R_1}/2\to\mu$. Besides, if the quantity $R$ from (\ref{B5}) is
equal to the root $R_1$, then it is easy to obtain asymptotic expansions at
$p_1\to\infty$ of quasiparticle energies,
\begin{eqnarray}
|p_{01}|&=&|p_1|-\mu+\mu_5+\frac{\Delta^2+M^2}{2|p_1|} +{\cal O}\big
(1/p_1^2\big ),~~ |\bar p_{01}|=|p_1|+\mu+\mu_5+\frac{\Delta^2+M^2}{2|p_1|}
+{\cal O}\big (1/p_1^2\big ),\nonumber\\
|p_{02}|&=&|p_1|+\mu-\mu_5+\frac{\Delta^2+M^2}{2|p_1|} +{\cal O}\big
(1/p_1^2\big ),~~ |\bar p_{02}|=|p_1|-\mu-\mu_5+\frac{\Delta^2+M^2}{2|p_1|}
+{\cal O}\big (1/p_1^2\big ),\label{B26}
\end{eqnarray}
which follow from (\ref{B701})--(\ref{B71}) as well as from the relations
(\ref{B5}) and (\ref{B41}). As a result, it can be established from (\ref{B26})
the asymptotic expansion (\ref{33}). We would like to emphasize once again that
the asymptotic behavior (\ref{33}) does not depend on which of the roots
$R_1$, $R_2$ or $R_3$ of the equation (\ref{B6}) is taken as the quantity $R$ from
the relations (\ref{B5}).

Now, taking into account the relations (\ref{B1}) and (\ref{B3}) one can
integrate over $p_0$ in (\ref{25}), using the well-known formula
\begin{equation}
\int_{-\infty}^\infty~dp_0\ln\left(p_0-A\right)=i\pi |A|\label{B15}
\end{equation}
(obtained rigorously, e.g., in Appendix B of \cite{gubina2} and true up to an
infinite term independent on the real quantity $A$). As a result, we have
\begin{eqnarray}
\frac{i}{2}\int\frac{d^2p}{(2\pi)^2}\ln\left[\det \bar
B(p)\right]&=&-\int_{-\infty}^\infty\frac{dp_1}{8\pi}\Big\{|p_{01}|+|p_{02}|+|\bar
p_{01}|+|\bar p_{02}|\Big\}\nonumber\\
&=&-\int_0^\infty\frac{dp_1}{4\pi}\Big\{|p_{01}|+|p_{02}|+|\bar p_{01}|+|\bar
p_{02}|\Big\},\label{B9}
\end{eqnarray}
where the expressions for energies of quasiparticles, $p_{01}$ and $p_{02}$,
as well as for energies of quasiantiparticles, $\bar p_{01}$ and $\bar
p_{02}$, are given in (\ref{B41}). Note that the last equality in (\ref{B9}) is due
to the fact that the sum $(|p_{01}|+|p_{02}|+|\bar p_{01}|+|\bar p_{02}|)$ is
an even function of $p_1$, as it is easily seen from (\ref{B41}) and (\ref{B5}).
Moreover, due to (\ref{B41}) one can obtain an equivalent expression for
(\ref{B9}),
\begin{eqnarray}
\frac{i}{2}\int\frac{d^2p}{(2\pi)^2}\ln\left[\det\bar
B(p)\right]&=&-\int_0^\infty\frac{dp_1}{4\pi}\Big\{\sqrt{r^2-4q}+\sqrt{r^2-4s}\nonumber\\&+&(r-\sqrt{r^2-4q})\theta
(r-\sqrt{r^2-4q}) +(r-\sqrt{r^2-4s})\theta (r-\sqrt{r^2-4s})\Big\},\label{B91}
\end{eqnarray}
where $r$, $q$ and $s$ are defined in (\ref{B3})--(\ref{B5}) and $\theta(x)$ is the
Heaviside step function. It is very convenient to use this relation in the
cases $r=\sqrt{R_1}$ and $r=\sqrt{R_2}$. In these cases an ultraviolet
divergence of the integral (\ref{B91}) is due to the first two terms in the
braces, whereas the terms with $\theta(x)$ functions do not generate any
divergences there.

Finally, in addition to (\ref{B42}), we would like to present two other
particular cases, in which the expressions for the quasiparticle energies
(\ref{B41}) and the sum $(|p_{01}|+|p_{02}|+|\bar p_{01}|+|\bar p_{02}|)$ can
be given in an exact analytical form. Namely, if $\Delta=0$ then we have from
(\ref{26}) that each of $p_{01}$, $p_{02}$, $\bar p_{01}$, and $\bar p_{02}$ is
given by one of the expressions $\mu\pm\sqrt{M^2+(p_1-\mu_5)^2}$ or
$-\mu\pm\sqrt{M^2+(p_1+\mu_5)^2}$. Therefore, their sum is represented by
\begin{eqnarray}
\big (|p_{01}|+|p_{02}|+|\bar p_{01}|+|\bar p_{02}|\big )\big |_{\Delta=0}&=&\nonumber\\
\sum_{\eta=\pm}&&\hspace{-4mm}\left (\left
|\mu+\eta\sqrt{M^2+(p_1-\eta\mu_5)^2}\right |+\left
|\mu+\eta\sqrt{M^2+(p_1+\eta\mu_5)^2}\right |\right ). \label{36}
\end{eqnarray}
Analogously, at $M=0$ we have from (\ref{27})
\begin{eqnarray}
\big (|p_{01}|+|p_{02}|+|\bar p_{01}|+|\bar p_{02}|\big )\big |_{M= 0}&=&\nonumber\\
\sum_{\eta=\pm}&&\hspace{-4mm}\left (\left
|\mu_5+\eta\sqrt{\Delta^2+(p_1-\eta\mu)^2}\right |+\left
|\mu_5+\eta\sqrt{\Delta^2+(p_1+\eta\mu)^2}\right |\right ). \label{37}
\end{eqnarray}

\section{Derivation of the relation (\ref{350})}
\label{ApD}

If $\Delta=0$ and $M\ne 0$, then the quasiparticle energies (\ref{B41}) can be
easily found from the expression (\ref{26}). In this case each of $p_{01}$,
$p_{02}$, $\bar p_{01}$, and $\bar p_{02}$ is given by one of the expressions
$\mu\pm\sqrt{M^2+(p_1-\mu_5)^2}$ or $-\mu\pm\sqrt{M^2+(p_1+\mu_5)^2}$;
therefore their sum is represented by the formula (\ref{36}). Taking this
relation into account as well as using the well-known relations $|x|=x\theta
(x)-x\theta (-x)$ and $\theta (x)=1-\theta(-x)$, it is possible to bring the
expression (\ref{35}) at $\Delta=0$ and $M\ne 0$ to the following form:
\begin{eqnarray}
\Omega^{ren} (M,\Delta=0)&=&-\frac{M^2}{4\pi}+\frac{M^2}{2\pi}\ln\left
(\frac{M}{M_1}\right )-U-V, \label{D1}
\end{eqnarray}
where
\begin{eqnarray}
U&=&\int_0^\infty\frac{dp_1}{2\pi}\Big\{\sqrt{M^2+(p_1+\mu_5)^2}+\sqrt{M^2+(p_1-\mu_5)^2}-2\sqrt{M^2+p_1^2}\Big\},\label{D2}\\
V&=&\int_0^\infty\frac{dp_1}{2\pi}\Big\{\Big (\mu-\sqrt{M^2+(p_1-\mu_5)^2}\Big
) \theta\Big (\mu-\sqrt{M^2+(p_1-\mu_5)^2}\Big )\nonumber\\&&~~~~~~~+\Big
(\mu-\sqrt{M^2+(p_1+\mu_5)^2}\Big ) \theta\Big
(\mu-\sqrt{M^2+(p_1+\mu_5)^2}\Big )\Big\}.\label{D3}
\end{eqnarray}
The convergent improper integral $U$ can be represented in the form
\begin{eqnarray}
U=\lim_{\Lambda\to\infty}\left\{\int_0^\Lambda\frac{dp_1}{2\pi}\sqrt{M^2+(p_1+\mu_5)^2}+\int_0^\Lambda\frac{dp_1}{2\pi}\sqrt{M^2+(p_1-\mu_5)^2}-2\int_0^\Lambda\frac{dp_1}{2\pi}\sqrt{M^2+p_1^2}\right\}.
\label{D4}
\end{eqnarray}
Denoting the first (second) integral in the braces of (\ref{D4}) as $U_1$ (as
$U_2$) and carrying out there the change of variables, $q=p_1+\mu_5$
($q=p_1-\mu_5$), one can obtain
\begin{eqnarray}
\label{D5} U_1&=&\int_{\mu_5}^{\Lambda+\mu_5}\frac{dq}{2\pi}
\sqrt{M^2+q^2}\equiv\left
(\int_{0}^\Lambda+\int_{\mu_5}^{0}+\int_{\Lambda}^{\Lambda+\mu_5}\right )\frac{dq}{2\pi}\sqrt{M^2+q^2},\\
U_2&=& \int_{-\mu_5}^{\Lambda-\mu_5}\frac{dq}{2\pi}\sqrt{M^2+q^2}\equiv \left
(\int_{0}^\Lambda+\int^{0}_{-\mu_5}+\int_{\Lambda}^{\Lambda-\mu_5}\right
)\frac{dq}{2\pi}\sqrt{M^2+q^2}. \label{D6}
\end{eqnarray}
Substituting (\ref{D5}) and (\ref{D6}) in (\ref{D4}) and taking into account
that
\begin{eqnarray}
\left (\int_{\mu_5}^{0}+\int^{0}_{-\mu_5}\right
)\frac{dq}{2\pi}\sqrt{M^2+q^2}=0, \label{D7}
\end{eqnarray}
we have after a direct integration
\begin{eqnarray}
U=\lim_{\Lambda\to\infty}\left
(\int_{\Lambda}^{\Lambda+\mu_5}-\int^{\Lambda}_{\Lambda-\mu_5}\right
)\frac{dq}{2\pi}\sqrt{M^2+q^2}=\frac{\mu_5^2}{2\pi}. \label{D8}
\end{eqnarray}
Analogously, the quantity $V$ (\ref{D3}) can be represented as a sum of two
integrals, in which one should perform a change of variables similar to $U_1$
and $U_2$, correspondingly. As a result, we have
\begin{eqnarray}
V&=&\int_{-\mu_5}^\infty\frac{dq}{2\pi}\Big (\mu-\sqrt{M^2+q^2}\Big )
\theta\Big (\mu-\sqrt{M^2+q^2}\Big )+\int_{\mu_5}^\infty\frac{dq}{2\pi}\Big
(\mu-\sqrt{M^2+q^2}\Big ) \theta\Big
(\mu-\sqrt{M^2+q^2}\Big )\nonumber\\
&=&\theta\Big (\mu-M\Big )\int_{0}^{\sqrt{\mu^2-M^2}}\frac{dq}{\pi}\Big
(\mu-\sqrt{M^2+q^2}\Big ).\label{D9}
\end{eqnarray}
Direct integration in (\ref{D9}) gives us the following expression for $V$:
\begin{eqnarray}
V&=&\frac{\theta (\mu-M)}{2\pi}\left
(\mu\sqrt{\mu^2-M^2}-M^2\ln\frac{\mu+\sqrt{\mu^2-M^2}}{M}\right ). \label{D10}
\end{eqnarray}
Hence, taking into account the relations (\ref{D10}), (\ref{D8}), and
(\ref{D1}), we are convinced of the validity of the formula (\ref{350}). By analogy,
one can derive the expression (\ref{035}).

\section{Summation over Matsubara frequencies}
\label{ApC}

Let us sum the series
\begin{eqnarray}
S(a)=\sum^{\infty}_{n=-\infty}\ln (i\omega_n-a),\label{C1}
\end{eqnarray}
where $\omega_n=\pi T(2n+1)$ and $a$, $T$ are some real quantities. The
expression can be modified in the following way:
\begin{eqnarray}
S(a)=\sum^{\infty}_{n=0}\left\{\ln (i\omega_n-a)+\ln
(-i\omega_n-a)\right\}=\sum^{\infty}_{n=0}\ln (a^2+\omega_n^2).\label{C2}
\end{eqnarray}
It is easy to find from (\ref{C2})
\begin{eqnarray}
\frac{dS(a)}{da}=2a\sum^{\infty}_{n=0}(a^2+\omega_n^2)^{-1}=\frac{\beta}2\tanh\left
(\frac{\beta a}{2}\right ),\label{C3}
\end{eqnarray}
where we have used the well-known relation
\begin{eqnarray*}
\sum^{\infty}_{n=0}(b^2+(2n+1)^2)^{-1}=\frac{\pi}{4b}\tanh\left (\frac{\pi
b}{2}\right ).
\end{eqnarray*}
Finally, integrating both sides of the relation (\ref{C3}) with respect to the
variable $a$ and omitting an unessential constant independent on the quantity $a$,
one can obtain
\begin{eqnarray}
S(a)=\ln\left [\exp (\beta a/2)+\exp (-\beta a/2)\right ]=\ln\left [\exp (\beta
|a|/2)+\exp (-\beta |a|/2)\right ]=\frac{\beta |a|}{2}+\ln\left [1+\exp (-\beta
|a|)\right ].\label{C4}
\end{eqnarray}


\begin{thebibliography}{999}
\bibitem{3+1}
D.V. Deryagin, D.Y. Grigoriev and V.A. Rubakov, Int. J. Mod. Phys. A {\bf 7},
659 (1992); M.~Sadzikowski and W.~Broniowski, Phys.\ Lett.\ B {\bf 488}, 63
(2000);
 W.~Broniowski,
  Acta Phys.\ Polon.\ Supp.\  {\bf 5}, 631 (2012).

\bibitem{nakano}
 E.~Nakano and T.~Tatsumi,
  Phys.\ Rev.\  D {\bf 71}, 114006 (2005).

\bibitem{Tatsumi:2014cea}
T.~Tatsumi and T.~Muto,
  arXiv:1403.1927 [nucl-th];
T.~Tatsumi, K.~Nishiyama and S.~Karasawa,
  arXiv:1405.2155 [hep-ph].

\bibitem{osipov}
  J.~Moreira, B.~Hiller, W.~Broniowski, A.A.~Osipov and A.H.~Blin,
  arXiv:1312.4942 [hep-ph].

\bibitem{nickel}
D.~Nickel,  Phys.\ Rev.\  D {\bf 80}, 074025 (2009);
S.~Carignano, D.~Nickel and M.~Buballa,
 Phys.\ Rev.\  D {\bf 82}, 054009 (2010);
 H.~Abuki, D.~Ishibashi and K.~Suzuki,
  arXiv:1109.1615.

\bibitem{maedan}
 S.~Maedan, Prog.\ Theor.\ Phys.\  {\bf 123}, 285 (2010);
 A.~Flachi,
  JHEP {\bf 1201}, 023 (2012);
arXiv:1304.6880 [hep-th].

\bibitem{Heinz:2013eu}
  A.~Heinz,
  arXiv:1301.3430 [hep-ph].

\bibitem{pisarski}
 T.~Kojo, Y.~Hidaka, L.~McLerran and R.D.~Pisarski,
  Nucl.\ Phys.\  A {\bf 843}, 37 (2010).

\bibitem{miransky}
 E.V.~Gorbar, M.~Hashimoto and V.A.~Miransky,
  Phys.\ Rev.\ Lett.\  {\bf 96}, 022005 (2006);
J.O.~Andersen and T.~Brauner,
  Phys.\ Rev.\  D {\bf 81}, 096004 (2010);
C.f.~Mu, L.y.~He and Y.x.~Liu,
  Phys.\ Rev.\  D {\bf 82}, 056006 (2010).

\bibitem{zfk}
I.E. Frolov, K.G. Klimenko and V.Ch. Zhukovsky,   Phys.\ Rev.\  D {\bf 82},
076002 (2010);
 Moscow Univ.\ Phys.\ Bull.\  {\bf 65}, 539 (2010).

\bibitem{incera}
E.J.~Ferrer, V.~de la Incera and A.~Sanchez,
  arXiv:1205.4492.

\bibitem{Anglani:2013gfu}
R.~Anglani, R.~Casalbuoni, M.~Ciminale, R.~Gatto, N.~Ippolito, M.~Mannarelli
and M.~Ruggieri,
  arXiv:1302.4264 [hep-ph].


\bibitem{buballa} 
  M.~Buballa and S.~Carignano,
  arXiv:1406.1367 [hep-ph].

\bibitem{caldas} 
  H.~Caldas,
  J.\ Stat.\ Mech.\  {\bf 1110}, P10005 (2011).
  [arXiv:1106.0948 [cond-mat.str-el]].
\bibitem{Roscher}D. Roscher, J. Braun, and J.E. Drut, Phys.\ Rev.\  A {\bf 89}, 063609 (2014).
 
\bibitem{thies}
V.~Schon and M.~Thies, 
 Phys.\ Rev.\  D {\bf 62}, 096002 (2000);
A.~Brzoska and M.~Thies,
  Phys.\ Rev.\  D {\bf 65}, 125001 (2002).

\bibitem{thies2}
O.~Schnetz, M.~Thies and K.~Urlichs,
  Annals Phys.\  {\bf 314}, 425 (2004);
C.~Boehmer and M.~Thies,
   Phys.\ Rev.\  D {\bf 80}, 125038 (2009);
J.~Hofmann,  Phys.\ Rev.\  D {\bf 82}, 125027 (2010).

\bibitem{basar}
G.~Basar, G.V.~Dunne and M.~Thies,
  Phys.\ Rev.\  D {\bf 79}, 105012 (2009).

\bibitem{gubina}
  D.~Ebert, N.V.~Gubina, K.G.~Klimenko, S.G.~Kurbanov, V.C.~Zhukovsky,
  Phys.\ Rev.\ D {\bf 84}, 025004 (2011).

\bibitem{gubina2}
  N.V.~Gubina, K.G.~Klimenko, S.G.~Kurbanov and V.C.~Zhukovsky,
  Phys.\ Rev.\ D {\bf 86}, 085011 (2012);
 Moscow Univ.\ Phys.\ Bull.\  {\bf 67}, 131 (2012).

\bibitem{chodos}
 A.~Chodos, H.~Minakata, F.~Cooper, A.~Singh, and W.~Mao,
  Phys. Rev. D {\bf 61}, 045011 (2000).

\bibitem{ff}
P. Fulde and R.A. Ferrel, Phys. Rev. {\bf 135}, A550 (1964).

\bibitem{andrianov}
 A.A.~Andrianov, D.~Espriu and X.~Planells,
  Eur.\ Phys.\ J.\ C {\bf 73}, 2294 (2013).

\bibitem{andrianov2}
 A.A.~Andrianov, D.~Espriu and X.~Planells,
 Eur.\ Phys.\ J.\ C {\bf 74}, 2776 (2014).

\bibitem{ruggieri}
R.~Gatto and M.~Ruggieri,
  Phys.\ Rev.\ D {\bf 85}, 054013 (2012);
M.~Ruggieri,
  arXiv:1110.4907.

\bibitem{huang}
  L.~Yu, H.~Liu and M.~Huang,
  arXiv:1404.6969 [hep-ph].

\bibitem{pauli}
W. Pauli, Nuovo Cimento, {\bf 6}, 204 (1957); F. Gursey, Nuovo Cimento, {\bf
7}, 411, (1957).

\bibitem{weinberg}
S. Weinberg, ``The quantum Theory of Field II'', Cambridge Univ. Press, Cambridge, England, 1996.

\bibitem{fujikawa}
K. Fujikawa,   Phys.\ Rev.\  D {\bf 21}, 2848 (1980).

\bibitem{kzz}
  K.G.~Klimenko, R.N.~Zhokhov and V.C.~Zhukovsky,
  Phys.\ Rev.\ D {\bf 86}, 105010 (2012).

\bibitem{thies1}
M.~Thies,
  Phys.\ Rev.\ D {\bf 68}, 047703 (2003).

\bibitem{Ojima:1977cg}
  I.~Ojima and R.~Fukuda,
  Prog.\ Theor.\ Phys.\  {\bf 57}, 1720 (1977).

\bibitem{Vasiliev:1995qp}
  A.N.~Vasiliev and G.Y.~Panasyuk,
  Theor.\ Math.\ Phys.\  {\bf 103}, 570 (1995)
  [Teor.\ Mat.\ Fiz.\  {\bf 103}, 295 (1995)].

\bibitem{Klimenko:1986uq}
 K.G.~Klimenko,
  Theor.\ Math.\ Phys.\  {\bf 75}, 487 (1988)
  [Teor.\ Mat.\ Fiz.\  {\bf 75}, 226 (1988)].

\bibitem{ohwa}
K.~Ohwa, Phys.\ Rev.\  D {\bf 65}, 085040 (2002).

\bibitem{Ebert:2013dda}
D.~Ebert, T.G.~Khunjua, K.G.~Klimenko and V.C.~Zhukovsky,
  Int.\ J.\ Mod.\ Phys.\ A {\bf 29}, 1450025 (2014);
 V.C.~Zhukovsky, K.G.~Klimenko and T.G.~Khunjua,
  Moscow Univ.\ Phys.\ Bull.\  {\bf 68}, 105 (2013)
  [Vestn.\ Mosk.\ Univ.\ Fiz.\ Astron.\  {\bf 2}, 11 (2013)].
\bibitem {VCHZH} A.A. Sokolov, I.M. Ternov, V.Ch. Zhukovsky, and
  A.V. Borisov, ``Kalibrovochnye Polya'' (``Gauge Fields'') (in Russian),
  Moscow University Publishing House, Moscow, 1985 
\bibitem{jacobs}
L. Jacobs, Phys.\ Rev.\  D {\bf 10}, 3956 (1974); K.G.~Klimenko, Theor.\ Math.\
Phys.\  {\bf 70}, 87 (1987).

\bibitem{vasiliev}
A.N. Vasiliev,``Functional methods in quantum field theory and statistical
physics'', Leningrad Univ. Press, Leningrag, USSR, 1976.

\bibitem{Ebert:2009ty}
  D.~Ebert, K.G.~Klimenko,  Phys.\ Rev.\  {\bf D80}, 125013 (2009).

\bibitem{Birkhoff}
G. Birkhoff and S. Mac Lane, ``A Survey of Modern Algebra``, New York:
Macmillan, 1977.

\end{thebibliography}
\end{document}